\pdfoutput=1

\documentclass[aps,twocolumn,amsmath,amssymb,preprintnumbers,floatfix]{revtex4-1}

\usepackage[utf8]{inputenc}
\usepackage{newtxtext}
\usepackage[upint]{newtxmath}
\usepackage{microtype}
\usepackage{textcomp}
\usepackage{eucal}
\usepackage{siunitx}
\usepackage{graphicx}
\usepackage{caption}
\usepackage{color}
\usepackage[caption=false]{subfig}
\usepackage{float}

\usepackage{enumerate}
\usepackage{amsfonts}
\usepackage{amsmath}
\usepackage{amssymb}

\usepackage{bm}
\usepackage{color}
\usepackage{soul}

\usepackage{graphicx}

\usepackage[colorlinks,allcolors=blue]{hyperref}
\usepackage[capitalize]{cleveref}

\let\phi=\varphi
\let\epsilon=\varepsilon

\definecolor{DarkRed}{rgb}{0.80,0,0}
\definecolor{lightBlue}{rgb}{0.5,0.63,0.9}
\definecolor{green}{rgb}{0,0.8,0.6}

\newcommand{\prlsection}[1]{\textit{#1}.\kern0.05em---\kern0.05em\ignorespaces}

\begin{document}
\title{The effect of midgap states on the magnetic exchange interaction mediated by a $d$-wave superconductor}
\author{Atousa Ghanbari}
\affiliation{\mbox{Center for Quantum Spintronics, Department of Physics, Norwegian University of Science and Technology,}\\NO-7491 Trondheim, Norway}

\author{Eirik Erlandsen}
\affiliation{\mbox{Center for Quantum Spintronics, Department of Physics, Norwegian University of Science and Technology,}\\NO-7491 Trondheim, Norway}

\author{Jacob Linder}
\affiliation{\mbox{Center for Quantum Spintronics, Department of Physics, Norwegian University of Science and Technology,}\\NO-7491 Trondheim, Norway}
\begin{abstract}
We theoretically study the indirect interaction between two ferromagnetic contacts located on the surface of a $d$-wave superconductor. When the magnets are connected to a $\{010\}$ edge of the superconductor we find an oscillating RKKY interaction that varies in sign as the distance between the magnetic contacts is varied. However, when coupling the magnets to an $\{110\}$ edge of the superconductor, we find that the presence of midgap states qualitatively changes the results. The ground state of the system is then found to always favor alignment of the magnets as this configuration most strongly suppresses the midgap states, leading to a larger condensation energy which dominates over the intrinsic RKKY interaction. 
\end{abstract}
\maketitle


\section{Introduction}
\begin{figure}[H]
\includegraphics[width=0.4\textwidth]{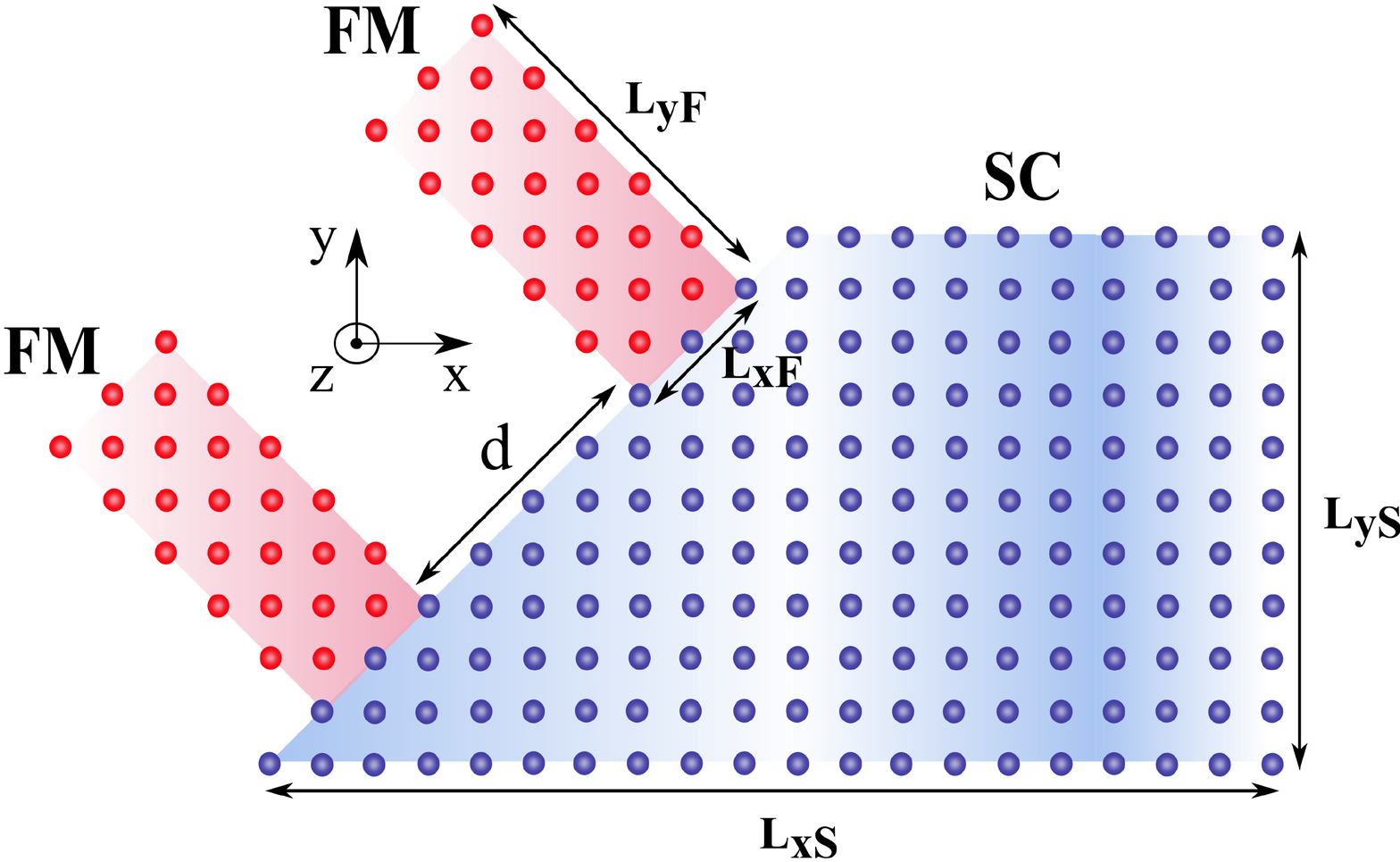}
\caption{Schematic illustration of a $d$-wave superconductor with midgap surface states mediating the indirect exchange interaction between two ferromagnetic contacts. We will consider configurations where the magnetization of the two magnets is either parallel (P) or antiparallel (AP). For comparison, the ferromagnetic contacts will also be attached to the lower, non-diagonal, edge of the superconductor. The lengths indicated on the figure will in the following be specified by the number of atomic distances.} 
\label{fig:model1}
\end{figure}

Superconductors with $d$-wave symmetry have an anisotropic order parameter which drops to zero along some nodal directions \cite{d_wave_1,d_wave_2,d-wave_3}. A $\{110\}$ edge of a $d_{x^2 - y^2}$ superconductor has been shown to feature dispersionless surface states with zero energy, called midgap states \cite{Midgap_states_Chia}. The appearance of midgap states for such an edge is related to the fact that the order parameter in a $45^{\circ}$ rotated coordinate system takes the form $d_{xy}$, introducing opposite signs for the pair potential experienced by particles undergoing specular and Andreev reflections at the surface. The $\{110\}$ edge also gives rise to a zero bias conductance peak \cite{ZBCP_theory_1}, which is a result of the presence of the midgap states  \cite{Midgap_states_and_ZBCP}. Such a zero bias conductance peak has been experimentally observed in the  high-$T_c$ cuprate superconductors \cite{ZBCP_exp_1, ZBCP_exp_2, ZBCP_exp_3,ZBCP_exp_4} and has been important in determining the pairing symmetry of these superconductors. \\ 
\indent The indirect exchange interaction between two localized spins, mediated by the itinerant electrons of a host material, was first introduced by Ruderman, Kittel, Kasuya and Yosida, and is known as the RKKY interaction \cite{RKKY_normal_metal1,RKKY_normal_metal2,RKKY_normal_metal3}. In this indirect exchange interaction, itinerant electrons of the host material scatter off a localized spin, and the wave functions of the scattered electrons interfere with each other giving rise to alternating regions with high density of spin up/down. This leads to the well-known RKKY oscillations in the spin-spin interaction strength which decrease with the distance $R$ between the two localized spins as $R^{- D}$, where $D$ is the dimensionality of the system. RKKY interaction has been investigated in various materials ranging from normal metals \cite{RKKY_normal_metal1, RKKY_normal_metal3}, to one- and two-dimensional electron gases \cite{RKKY_1D_electron_gas,RKKY_1or2D_EG_Imamura}, two-dimensional structures like graphene \cite{Graphene1,Graphene2,Graphene3,Graphene4,Graphene5}, and topological insulators \cite{RKKY_topological_insulator1,RKKY_topological_insulator2,RKKY_topological_insulator3}.\\
\indent For a system consisting of magnetic impurities embedded in a superconductor, the influence of superconductivity on the indirect impurity-impurity interaction has also been studied in the literature \cite{RKKY_SC_2, RKKY_SC_5}. For a conventional $s$-wave superconductor, when the distance between the impurities is larger than the superconducting coherence length, the interaction between them is found to be antiferromagnetic in character and suppressed compared to the normal metal case. The suppression is caused by the superconducting gap reducing the number of states close to the Fermi level that can mediate the interaction. Below the coherence length, the behavior is similar to the normal metal case with an oscillatory RKKY interaction that changes sign with distance. However, non-perturbative treatments have shown that Yu-Shiba-Rusinov (YSR) bound states can give rise to mainly antiferromagnetic behavior even at distances shorter than the coherence length \cite{RKKY_SC_3}. Further, for impurities on the surface of  a three-dimensional topological insulator with proximity-induced s-wave superconductivity, the RKKY interaction favors the impurity spins to be in-plane and antiparallel \cite{RKKY_SC_4}. For a spin-valve structure consisting of two ferromagnetic insulators connected by an $s$-wave superconductor, experiments have shown that anti-alignment of the magnets is still favored \cite{RKKY_SC_6}.\\
\indent Conventional $s$-wave superconductors do however typically have coherence lengths far exceeding the decay length of the RKKY interaction. On the other hand, $d$-wave superconductors can feature very short coherence lengths of the order of nanometers \cite{YBCO_2}, offering an intriguing platform for studying the interplay between superconductivity and RKKY interaction, as the characteristic length scales of both phenomena are comparable. RKKY interaction between magnetic impurities mediated by a d-wave superconductor with an anisotropic order parameter of the type $d_{x^2 -y^2}$ has lead to similar behavior as in the s-wave case \cite{RKKY_SC_1}. Further, for a spin-valve structure involving a $d_{x^2 - y^2}$ superconductor, nodal quasiparticles close to the Fermi surface have been observed to mediate interaction that favors anti-alignment of the magnetic insulators for a sufficiently large superconductor thickness \cite{RKKY_SC_7}.\\
\indent As the gapped band structure of a superconductor suppresses the RKKY interaction, it is of interest to investigate the effect the presence of midgap states have on the interaction. We therefore consider a $d_{x^2 - y^2}$ superconductor and calculate the exchange interaction between two ferromagnetic contacts located on a diagonal $\{110\}$ edge, as illustrated in Fig.\! \ref{fig:model1}. The superconductor is modelled by an extended BCS tight binding Hamiltonian on a square lattice and connected to the metallic magnets through a hopping term across the interface. The results are obtained through a self-consistent solution of the Bogoliubov-de Gennes (BdG) equations \cite{deGennes1999}. To put the results into context, we consider the cases of a normal metal and an isotropic $s$-wave superconductor, in addition to the $d$-wave superconductor. In all three cases, we investigate the interaction between ferromagnetic contacts located on both diagonal and
\noindent non-diagonal edges.\\
\indent For magnetic contacts located on a diagonal edge of a $d$-wave superconductor, we find that the system always favors alignment of the two magnets. The variation in the strength of the magnetic exchange interaction as we vary the distance between the magnets is small compared to the magnitude of the interaction itself. We attribute these results to the aligned magnets more
\noindent efficiently suppressing the midgap states than the anti-aligned configuration. Although the aligned magnets induce a stronger  spin-splitting in the superconductor, suppressing the gap, the reduction of the midgap states leads to an overall larger gap and increased condensation energy. The parallel magnet configuration is therefore the ground state of the system.\\
\indent The paper is organized as follows. In Sec.\! \ref{sec:model} we introduce the model and methodology. Then, in Sec.\! \ref{sec:results} we present and discuss the results. Finally, in Sec.\! \ref{sec:summary} we provide a summary of the findings. The phase diagram of our $d$-wave superconductor model for a square system with continuous boundary conditions is included in the Appendix.

\section{Model and Methods}
\label{sec:model}
By means of a tight binding Hamiltonian on a square lattice, we model the attractive electron-electron interaction in a superconductor
\begin{align} \label{eq:eq1}
\begin{aligned}
H^{SC}= & -\sum_{\langle i,j \rangle,\alpha}t_{ij}c_{i\alpha}^{\dagger}c_{j\alpha}-\sum_{i,\alpha} \mu_{i} n_{i\alpha} -\sum_i U_i n_{i\uparrow} n_{i\downarrow}\\
&+\!\sum_{\langle ij\rangle, \alpha \neq \alpha^{\prime}} V_{ij} n_{i\alpha} n_{j\alpha^{\prime}}+\!\sum_{\langle ij\rangle , \alpha} V_{ij}^{\prime} n_{i\alpha} n_{j\alpha}.
\end{aligned}
\end{align}
Here, $c_{i\alpha}^{\dagger}$ is a creation operator creating an electron with spin $\alpha$ on lattice site $i=(i_x,i_y)$. The hopping amplitude is denoted by $t_{ij}$, and $\mu_i$ is the chemical potential. The third term represents on-site attractive interactions between opposite spins, where the number operator is $n_{i\alpha}=c_{i\alpha}^{\dagger}c_{i\alpha}$. This term gives rise to conventional spin singlet isotropic $s$-wave superconductivity. The fourth and fifth terms represent nearest neighbour interaction between opposite or equal spins, respectively. These terms can give rise to $d$-wave, $p$-wave or extended s-wave pairing for an attractive interaction potential. For the purposes of this article, we will set $V_{ij}^{\prime}$ to zero as we will not be interested in the possibility of equal spin pairing. As shown in the Appendix, the above model without $V_{ij}^{\prime}$ can give rise to a $d$-wave superconductor for a suitable choice of chemical potential. 
\subsection{Analytical Methods}
Through a mean-field treatment, we simplify the interaction terms. The on-site part of the interaction becomes
\begin{align}\label{eq:eq2}
-\sum_i U_i n_{i\uparrow} n_{i\downarrow}=-\sum_i U_i & \Big(c_{i\uparrow}^{\dagger} c_{i\downarrow}^{\dagger} \langle c_{i\downarrow} c_{i\uparrow}\rangle+c_{i\downarrow} c_{i\uparrow} \langle c_{i\uparrow}^{\dagger} c_{i\downarrow}^{\dagger}\rangle \notag\\
&-\langle c_{i\downarrow} c_{i\uparrow}\rangle \langle c_{i\uparrow}^{\dagger} c_{i\downarrow}^{\dagger}\rangle\Big).
\end{align}    
Defining the superconducting gap for the on-site interaction as $\Delta_i=-U_i\langle c_{i\downarrow} c_{i\uparrow}\rangle$, we obtain
\begin{equation}\label{eq:eq3}
-\sum_i U_i n_{i\uparrow} n_{i\downarrow}=\sum_i (c_{i\uparrow}^{\dagger} c_{i\downarrow}^{\dagger} \Delta_i+c_{i\downarrow} c_{i\uparrow} \Delta_i^{\ast})+H_0^S\,,      
\end{equation}
where we have defined
\begin{equation}\label{eq:eq4}
H_0^{S}=\sum_i\frac{|\Delta_i|^2}{U_i}\,.
\end{equation}
The on-site interaction $U_i$ will be taken to a constant $U \geq 0$ in the superconductor, and zero elsewhere. Once again, performing a mean-field treatment, the attractive nearest neighbor interaction term becomes  
\begin{align}
\begin{aligned}
\label{eq:eq5}
\sum_{\langle ij\rangle, \alpha \neq \alpha^{\prime}} &V_{ij} n_{i\alpha} n_{j\alpha^{\prime}} = \sum_{\langle ij\rangle, \alpha \neq \alpha^{\prime}}V_{ij}(c_{j\alpha^{\prime}}^{\dagger}c_{i\alpha}^{\dagger}\langle c_{i\alpha}c_{j\alpha^{\prime}} \rangle\\
&+c_{i\alpha}c_{j\alpha^{\prime}}\langle c_{j\alpha^{\prime}}^{\dagger}c_{i\alpha}^{\dagger} \rangle -\langle c_{j\alpha^{\prime}}^{\dagger}c_{i\alpha}^{\dagger} \rangle \langle c_{i\alpha}c_{j\alpha^{\prime}} \rangle)
.
\end{aligned}
\end{align}
We then define the nearest neighbor pairing amplitude
\begin{equation}\label{eq:eq6}
F_{ij}^{\alpha \alpha^{\prime}}=\langle c_{i\alpha}c_{j\alpha^{\prime}} \rangle,    
\end{equation}
transforming Eq.\! \eqref{eq:eq5} into
\begin{align}\label{eq:eq7}
&\sum_{\langle ij\rangle, \alpha \neq \alpha^{\prime}}V_{ij}(c_{j\alpha^{\prime}}^{\dagger}c_{i\alpha}^{\dagger}F_{ij}^{\alpha \alpha^{\prime}}+c_{i\alpha}c_{j\alpha^{\prime}}(F_{ij}^{\alpha \alpha^{\prime}})^\dagger) + H^{d}_0,
\end{align}
where
\begin{equation}\label{eq:eq8}
H_0^d=-\!\sum_{\langle i,j \rangle,\alpha \neq \alpha^{\prime}} V_{ij} |F_{ij}^{\alpha \alpha^{\prime}}|^2.    
\end{equation}
As $\sum_{\langle i,j \rangle} V_{ij} |F_{ij}^{\uparrow \downarrow}|^2 = \sum_{\langle i,j \rangle} V_{ji} |F_{ij}^{\downarrow \uparrow}|^2$, we can rewrite $H_0^d = -\sum_{\langle i,j \rangle} |F_{ij}^{\uparrow \downarrow}|^2 (V_{ij}  + V_{ji} )$, and similarly  
\begin{align}
\begin{aligned}
\label{eq:eq9}
\sum_{\langle ij\rangle, \alpha \neq \alpha^{\prime}}&\!V_{ij}(c_{j\alpha^{\prime}}^{\dagger}c_{i\alpha}^{\dagger}F_{ij}^{\alpha \alpha^{\prime}}+c_{i\alpha}c_{j\alpha^{\prime}}(F_{ij}^{\alpha \alpha^{\prime}})^\dagger)=  \\
 \sum_{\langle ij\rangle} &(c_{j\downarrow}^{\dagger}c_{i\uparrow}^{\dagger}F_{ij}^{\uparrow \downarrow}+c_{i\uparrow}c_{j\downarrow}(F_{ij}^{\uparrow \downarrow})^\dagger) (V_{ij} + V_{ji}).
\end{aligned}
\end{align}
In the following, we will take the nearest neighbor interaction to be $V_{ji} = V_{ij} = V \leq 0$ (corresponding to an attractive interaction) in the superconductor and zero elsewhere. The mean-field extended tight binding Hamiltonian now takes the form 
\begin{align}
\begin{aligned}
\label{eq:eq10}
&H_{mf}^{SC}= H_0 -\!\sum_{\langle i,j \rangle,\alpha}t_{ij}c_{i\alpha}^{\dagger}c_{j\alpha}-\sum_{i,\alpha} \mu_{i} n_{i\alpha}+\\
&\sum_i (c_{i\uparrow}^{\dagger} c_{i\downarrow}^{\dagger} \Delta_i+c_{i\downarrow} c_{i\uparrow} \Delta_i^{\ast})+\\
&2 \sum_{\langle ij\rangle} V (c_{j\downarrow}^{\dagger}c_{i\uparrow}^{\dagger}F_{ij}^{\uparrow \downarrow}+c_{i\uparrow}c_{j\downarrow}(F_{ij}^{\uparrow \downarrow})^\dagger),  
\end{aligned}
\end{align}
where $H_0 = H_0^S + H_0^d$.
\\
\indent The metallic ferromagnets that are attached to the superconductor are described by the following tight binding Hamiltonian
\begin{align} \label{eq:eq11}
H^{FM} = & -\sum_{\langle i,j \rangle,\alpha}t_{ij}c_{i\alpha}^{\dagger}c_{j\alpha}-\sum_{i,\alpha} \mu_{i} n_{i\alpha} \notag\\
&-\sum_{i\delta\alpha \beta}h_{i}^{\delta} (\sigma_z)_{\alpha \beta}\,c_{i\alpha}^{\dagger} c_{i\beta}\,.
\end{align}
The last term represents the coupling between the spin of an electron at site $i$ and the local magnetic exchange field, giving rise to ferromagnetism. The local exchange field $h^{\delta}_i$ is taken to produce a spin-splitting in the $z$-direction in spin-space, giving rise to a magnetization that could in general be either in-plane or out-of-plane in real space. Our model does not separate these cases as the magnetism is simply introduced through a spin-splitting. Orbital effects on the superconductor arising from the magnets, not considered in this model, can be limited by keeping the magnetization in-plane \cite{Bergeret2018}. The Pauli matrices are denoted by $\sigma$, and the index $\delta$ separates the local exchange field of each of the two magnets with $\delta = L,R$ for the leftmost and rightmost magnet, respectively. The sign of the local exchange field can be either the same or opposite for the two magnets, giving rise to parallel (P) or antiparallel (AP) ferromagnets. Outside of the magnets, the local magnetic exchange field is set to zero. The coupling between the magnets and the superconductor is introduced by having a nonzero hopping amplitude $t_{ij}$ across the ferromagnet-superconductor interfaces. The region outside of the superconductor and magnets is considered to be vacuum and decoupled from the rest of the system with a vanishing hopping amplitude. \\
\indent After diagonalization, the free energy of the system will be expressed as

\begin{equation}\label{eq:eq12}
F = H_0 - \frac{1}{2} \sum_{n=1}^{2N} E_n - \frac{1}{\beta} \sum _{n=1}^{2N} \text{ln} (1 + e^{-\beta E_n}),
\end{equation}
where $E_n$ is the quasiparticle energy associated with quantum number $n$, and $N$ is the number of lattice sites. The magnetic exchange interaction is computed as the difference in free energy between the configurations with parallel and antiparallel magnets 

\begin{align}
    J=F^{\uparrow \uparrow} - F^{\uparrow \downarrow},
\end{align}
\noindent which includes both the RKKY interaction mediated by the quasiparticles as well as the effect of the magnetic configurations on the condensation energy of the superconductor.

The Hamiltonian $H=H^{SC}+H^{FM}$ is diagonalized by means of the BdG method in order to compute the eigenvalues $E_n$ and eigenstates $\gamma_n$. The diagonalized Hamiltonian will then take the form 
\begin{align} \label{diag_H_3}
\hspace{-2cm}
    \begin{aligned}
           H & = & H_0 - \frac{1}{2} \sum_{n=1}^{2N} E_n + \sum_{n=1}^{2N} E_n \gamma_n^{\dagger} \gamma_n. 
    \end{aligned}
\end{align}
In order to perform the diagonalization, we start by rewriting the Hamiltonian as $ H=H_0+\frac{1}{2}\sum_{ij} B_i^{\dagger} h_{ij} B_j$ where we have introduced the basis
\begin{equation}\label{eq:eq13}
   B_i^{\dagger}=
  \left[ {\begin{array}{cccc}
    c_{i\uparrow}^{\dagger} & c_{i\downarrow}^{\dagger} & c_{i\uparrow} & c_{i\downarrow} 
  \end{array} } \right].
\end{equation}
Here, $h_{ij}$ is a $4 \times 4$ matrix that takes the following form for $i \neq j$  
\begin{equation} \label{eq:eq14}
   h_{ij}=
  \left[ {\begin{array}{cccc}
    -t & 0 & 0 & -2VF_{ij} \\

    0 & -t & 2VF_{ji} & 0 \\

    0 & 2V(F_{ij})^{\ast} & +t & 0\\

    -2V(F_{ji})^{\ast} & 0 & 0 & +t\\
  \end{array} } \right],
\end{equation}
and for $i=j$ 

\begin{align}\label{eq:eq15}
\begin{aligned}
   h_{ij}= 
  &\left[ {\begin{array}{cccc}
    -\mu_i+\sum_{\delta}h_i^{\delta} & 0 & 0 & \Delta_i \\
    0 & -\mu_i-\sum_{\delta}h_i^{\delta} & -\Delta_i & 0 \\
    0 & -(\Delta_i)^{\ast} & +\mu_i-\sum_{\delta}h_i^{\delta} & 0\\
    (\Delta_i)^{\ast} & 0 & 0 & +\mu_i+\sum_{\delta}h_i^{\delta}\\
  \end{array} } \right].
\end{aligned}
\end{align}
Writing the Hamiltonian on matrix form $H = H_0 + \frac{1}{2} W^{\dagger} S W$, and introducing the matrix P, we diagonalize the Hamiltonian $H = H_0 + \frac{1}{2} W^{\dagger} P^{\dagger} P S P^{\dagger} P W = H_0 + \frac{1}{2}  \tilde{W}^{\dagger} S_d \tilde{W}$. The eigenvectors of $S$ are

\begin{align}
  \begin{aligned}
       \Phi_n^{\dagger}= &
       \left[ {\begin{array}{ccccc}
       \phi_{1n}^{\ast} &  \cdots & \phi_{in}^{\ast} & \cdots & \phi_{Nn}^{\ast} 
       \end{array} } \right], 
  \\
     \varphi_{in}^{\ast} = &
  \left[ {\begin{array}{cccc}
    \upsilon_{in}^{\ast} & \nu_{in}^{\ast} & \omega_{in}^{\ast} & \chi_{in}^{\ast} 
  \end{array} } \right], 
  \end{aligned}
\end{align}
such that

\begin{equation}
   P^{\dagger} =  
     \left[ {\begin{array}{cccc}
   \Phi_1 & \Phi_2 & \dots & \Phi_{4N}
  \end{array} } \right]. 
\end{equation}
We next use $P^{\dagger} \tilde{W}= W$ along with

\begin{equation}
  \tilde{W}^{\dagger}=
  \left[ {\begin{array}{ccc}
    \gamma_{1}^{\dagger} & \dots & \gamma^{\dagger}_{4N}\\
  \end{array} } \right],
\end{equation}
and the relations between the quasiparticle operators that are not independent of each other. There are then $2N$ remaining independent quasiparticle operators with corresponding eigenvalues. The creation and annihilation operators $\{c^\dag,c\}$ can then be expressed in terms of quasiparticle creation and annihilation operators $\{\gamma^\dag,\gamma\}$  

\begin{equation}\label{eq:eq16}
\begin{array}{l@{\qquad}l}
\displaystyle c_{i\uparrow} = \sum_{n=1}^{2N} \upsilon_{i,n} \gamma_n + \omega_{i,n}^{\ast} \gamma_n^{\dagger} \,,
&
\displaystyle c_{i\downarrow} = \sum_{n=1}^{2N} \nu_{i,n} \gamma_n +  \chi_{i,n}^{\ast} \gamma_n^{\dagger}\,,
\\\\
\displaystyle c_{i\uparrow}^{\dagger} = \sum_{n=1}^{2N} \omega_{i,n} \gamma_n + \upsilon_{i,n}^{\ast} \gamma_n^{\dagger}\,,
&
\displaystyle c_{i\downarrow}^{\dagger} = \sum_{n=1}^{2N} \chi_{i,n} \gamma_n + \nu_{i,n}^{\ast} \gamma_n^{\dagger}\,.
\end{array}
\end{equation}
\indent Inserting these relations into the definition of the gap for the on-site interaction, we obtain the self-consistent gap equation 

\begin{equation}\label{eq:eq17}
\Delta_i = -U_i \sum_{n=1}^{2N} \Big [ ( \chi_{i,n}^{\ast} \upsilon_{i,n} - \nu_{i,n} \omega_{i,n}^{\ast} )  f(E_n) + \nu_{i,n} \omega_{i,n}^{\ast} \Big ].
\end{equation}
For the nearest neighbor pairing amplitudes, we introduce a simplified notation $F_{ij}^{\uparrow \downarrow} = F_{ij}$. Further, $F_{i,i+\hat{x}}$ is expressed as $F_i^{\hat{x}+}$ and likewise $F_{i+\hat{x},i} \equiv F_i^{+\hat{x}}$ and so on. Inserting the expressions from Eq.\! \eqref{eq:eq16} into the definitions of the pairing amplitudes, we obtain 

\begin{align} \label{eq:eq18}
\begin{aligned}
F_i^{x \pm} = & \sum_{n=1}^{2N} \Big[ (\omega_{i,n}^{\ast} \nu_{i \pm \hat{x},n}  - \upsilon_{i,n} \chi_{i \pm \hat{x},n}^{\ast}) f(E_n) + \upsilon_{i,n} \chi_{i \pm \hat{x},n}^{\ast} \Big ],
\\
F_i^{\pm x} = & \sum_{n=1}^{2N} \Big [ (\omega_{i \pm \hat{x},n}^{\ast} \nu_{i,n} - \upsilon_{i \pm \hat{x},n} \chi_{i,n}^{\ast}) f(E_n) + \upsilon_{i \pm \hat{x},n} \chi_{i,n}^{\ast} \Big ],
\\
F_i^{y \pm} = & \sum_{n=1}^{2N} \Big[ (\omega_{i,n}^{\ast} \nu_{i \pm \hat{y},n} - \upsilon_{i,n} \chi_{i \pm \hat{y},n}^{\ast}) f(E_n) + \upsilon_{i,n} \chi_{i \pm \hat{y},n}^{\ast} \Big ],
\\
F_i^{ \pm y} = & \sum_{n=1}^{2N} \Big [ (\omega_{i \pm \hat{y},n}^{\ast} \nu_{i,n}  - \upsilon_{i \pm \hat{y},n} \chi_{i,n}^{\ast}) f(E_n) + \upsilon_{i \pm \hat{y},n} \chi_{i,n}^{\ast} \Big ].
\end{aligned}
\end{align}

As we are interested in the effect of the midgap states on the indirect interaction between two ferromagnetic leads connected to the superconductor, establishing the presence of midgap states is of importance. This can be achieved by calculating the single particle local density of states (LDOS) which should have a peak around zero energy in the presence of midgap states. The number of charges on lattice site $i$ is given by $\rho_i=\sum_{\alpha}\langle c_{i\alpha}^\dagger c_{i\alpha} \rangle$, but this quantity can also be expressed as $\rho_i=\int\limits_{-\infty}^{+\infty}\!\! N_i(E)f(E)\textrm{d}E$. Here $N_i(E)$ is the local density of states at site $i$, and $f(E)$ is the Fermi-Dirac distribution with energy $E$ measured relative to the chemical potential. At $T=0$, we have $f(E)=1$ for $E < 0$ and $f(E)=0$ when $E > 0$. Comparing the two above expressions for the number of charges on lattice site $i$, the LDOS can then be expressed as
\begin{align} \label{eq:eq19}
    \begin{aligned}
       N_i(E) = \sum_{n=1}^{2N} \Big[ &\big(|\omega_{i,n}|^2 + |\chi_{i,n}|^2\big)\, \delta\big(E + E_n \big)\\
       +\,  &\big(|\upsilon_{i,n}|^2 + |\nu_{i,n}|^2\big) \,\delta\big(E - E_n \big)\Big]. 
    \end{aligned}
\end{align}
\indent Another quantity of interest is the magnetization on lattice site $i$, $\boldsymbol{M}_i = \langle \boldsymbol{S}_i \rangle$. Here, the spin operator is defined as $\boldsymbol{S}_i= \sum_{\alpha \beta} c_{i \alpha}^{\dagger} \boldsymbol{\sigma}_{\alpha \beta} c_{i \beta}$. The magnetization in the $z$-direction can then be expressed as 

\begin{align}
    \begin{aligned}
           M_i^z = \sum_{n=1}^{2N}\Big[ &\big( |\upsilon_{i,n}|^2 + |\chi_{i,n}|^2 - |\omega_{i,n}|^2 - |\nu_{i,n}|^2  \big)f\big(E_n\big) \\
           &+ |\omega_{i,n}|^2 - |\chi_{i,n}|^2\Big].
    \end{aligned}
\end{align}

\subsection{Computational Methods}

The computational part of this study consists of numerically diagonalizing the Hamiltonian and self-consistently solving the equations for either the on-site superconducting gap (Eq.\! \eqref{eq:eq17}) or the nearest neighbor pairing amplitudes (Eq.\! \eqref{eq:eq18}), depending on whether the superconductor is taken to be of the isotropic $s$-wave type or the $d$-wave type. Iterative solution of these equations require an initial value for the gap function/pairing amplitudes, and a convergence criterion in order to determine when a solution has been obtained. In this work, the convergence criterion was that the relative change in the gap/pairing amplitudes from one iteration to the next should be less than $1\times10^{-4}$ for the $d$-wave and $1\times10^{-3}$ for the $s$-wave state. The initial values for the $d$-wave state are listed in the Appendix and the initial value for the $s$-wave gap $\Delta$ was set to $0.5t$.

\begin{figure*}[t]
\subfloat{%
\hspace{-1.5cm}
\includegraphics[width=0.9\columnwidth,trim= 0.2cm 0.01cm 1.0cm 0.01cm,clip=true]{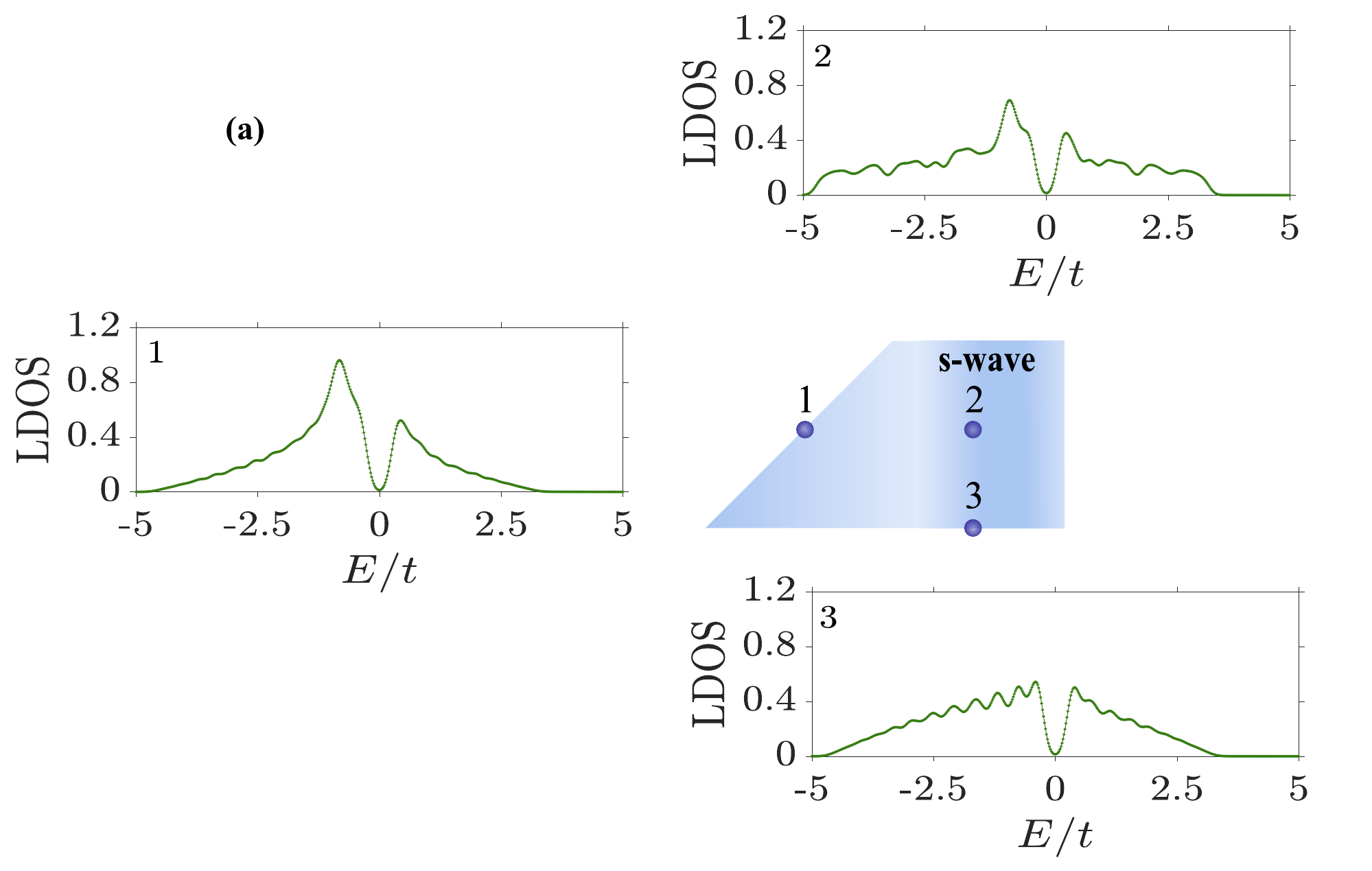}
}
\subfloat{%
\hspace{1cm}
\includegraphics[width=0.9\columnwidth,trim= 0.2cm 0.01cm 1.0cm 0.01cm,clip=true]{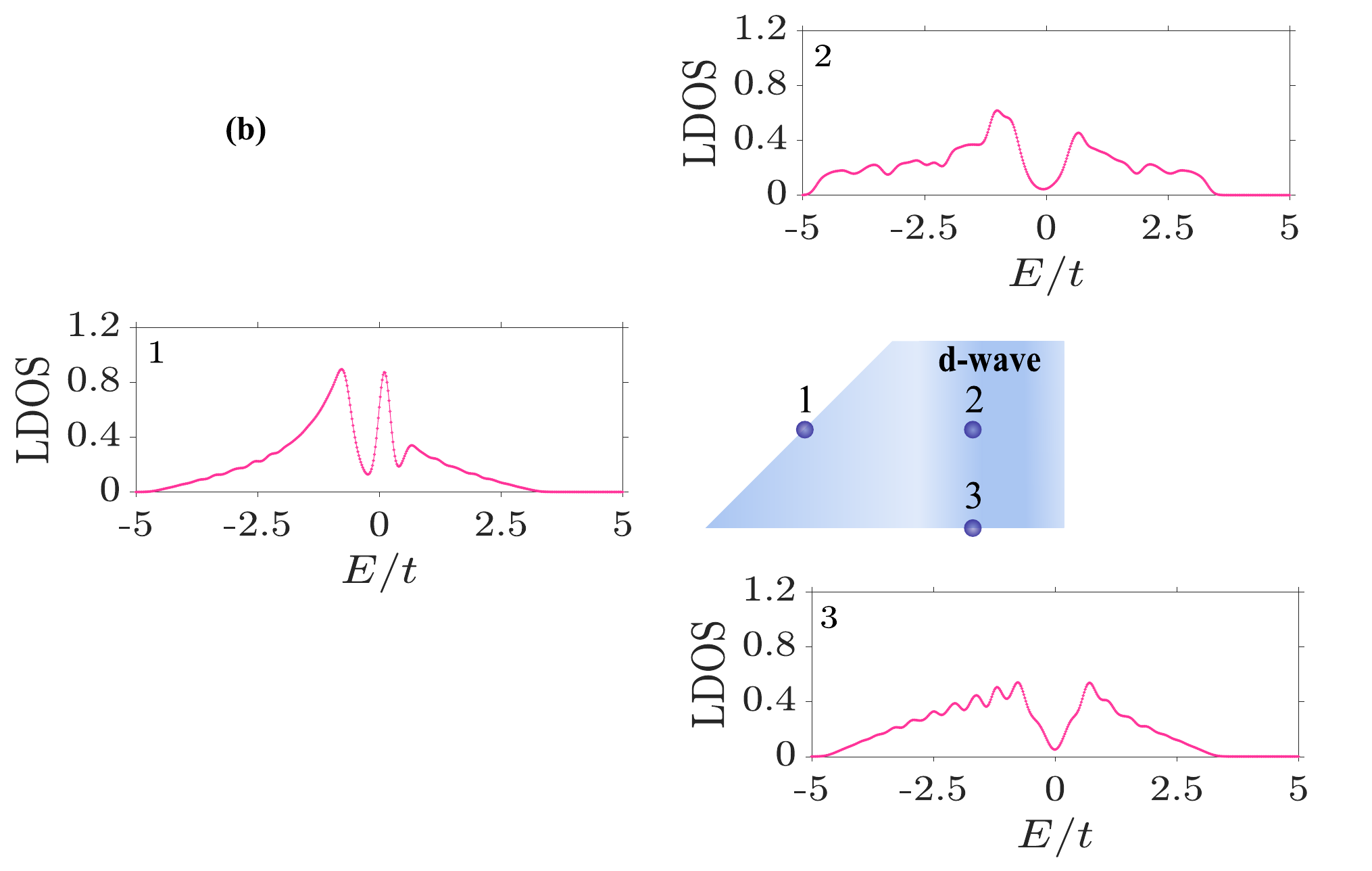}%
}
\caption{Local density of states (LDOS) for different points of an $s$-wave superconductor (a) and a $d$-wave superconductor (b), showing the presence of midgap states on the diagonal edge of the $d$-wave superconductor. In both cases the size of the structure is $L_{xS} = 34$ and $L_{yS} = 30$. For the $s$-wave results we have taken $U/t = 2$ and $V = 0$, while for the $d$-wave results we have taken $U = 0$ and $V/t$ = -1. In both cases we have set the chemical potential $\mu_S = 0.7t$.}
\label{fig:Midgap-states}
\end{figure*}

\section{Results and Discussion}
\label{sec:results}

We first investigate the presence of midgap surface states, i.e.\! zero energy states existing on an edge of a superconductor. As displayed in Fig.\! \ref{fig:Midgap-states}, we calculate the LDOS for different points on an $s$-wave and a $d$-wave superconductor without magnetic contacts. One of the points is located at the diagonal edge, one of the points is in the bulk, and the third point is on the lower horizontal edge. Only on the diagonal edge of the $d$-wave superconductor, Fig.\! \ref{fig:Midgap-states} (b1), there is a peak around zero energy signaling the presence of midgap states. In this figure, the chemical potential has been set to $\mu_S = 0.7t$, which gives rise to an asymmetric density of states around $E=0$ for our tight binding model as the gap in the electron spectrum is opened away from the middle of the band.\\
\indent As the presence of midgap states has been established, we move on to results for the indirect interaction between magnetic leads attached to a normal metal, an $s$-wave superconductor, and finally a $d$-wave superconductor. 

\begin{figure}[H]
\subfloat{%
\includegraphics[width=0.65\columnwidth,trim= 0.2cm 0.01cm 0.7cm 0.001cm,clip=true]{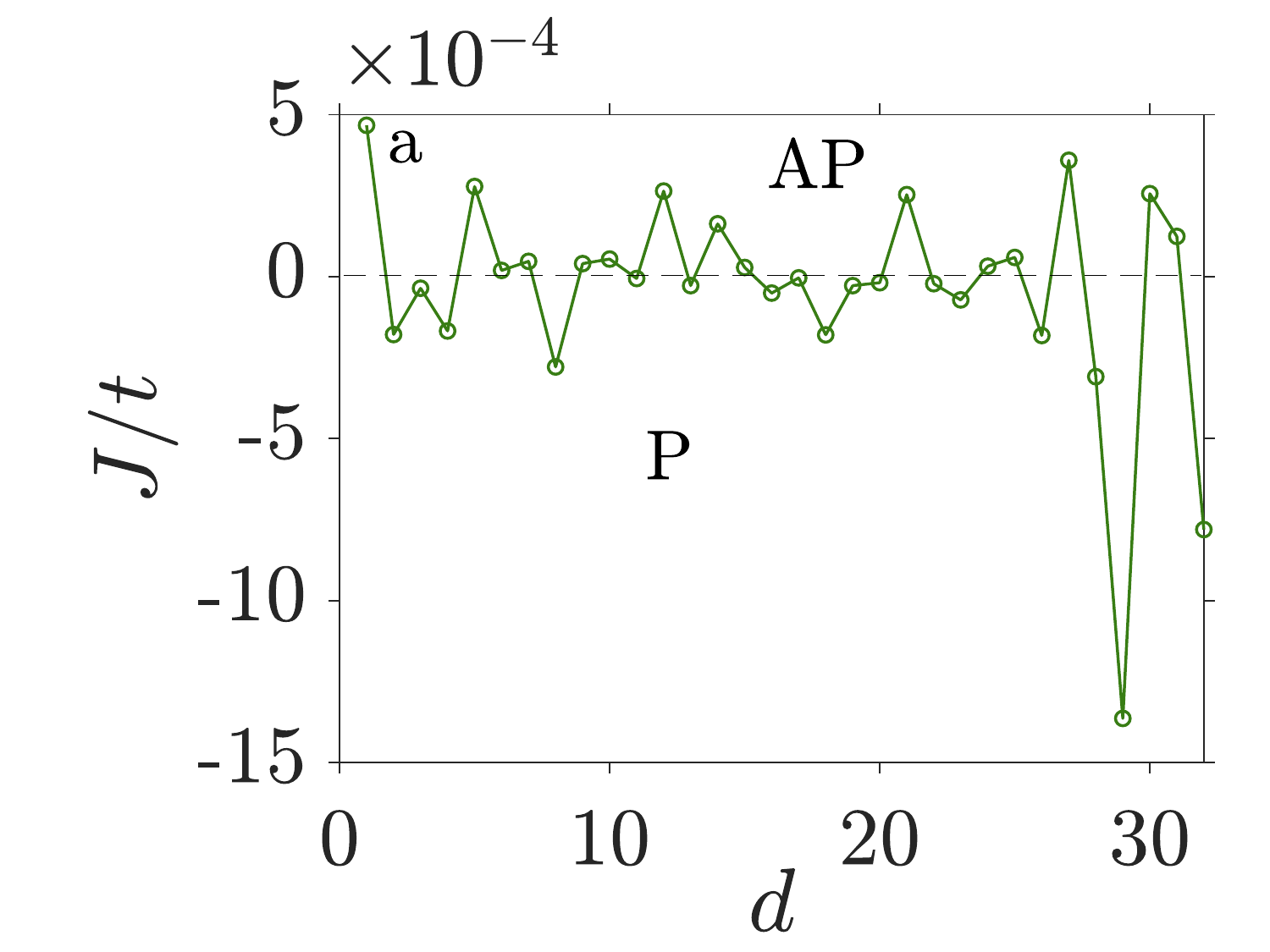}%
}
\subfloat{%
\hspace{0.1cm}
\includegraphics[width=0.35\columnwidth,trim= 0.02cm -14cm 1.0cm 0.01cm,clip=true]{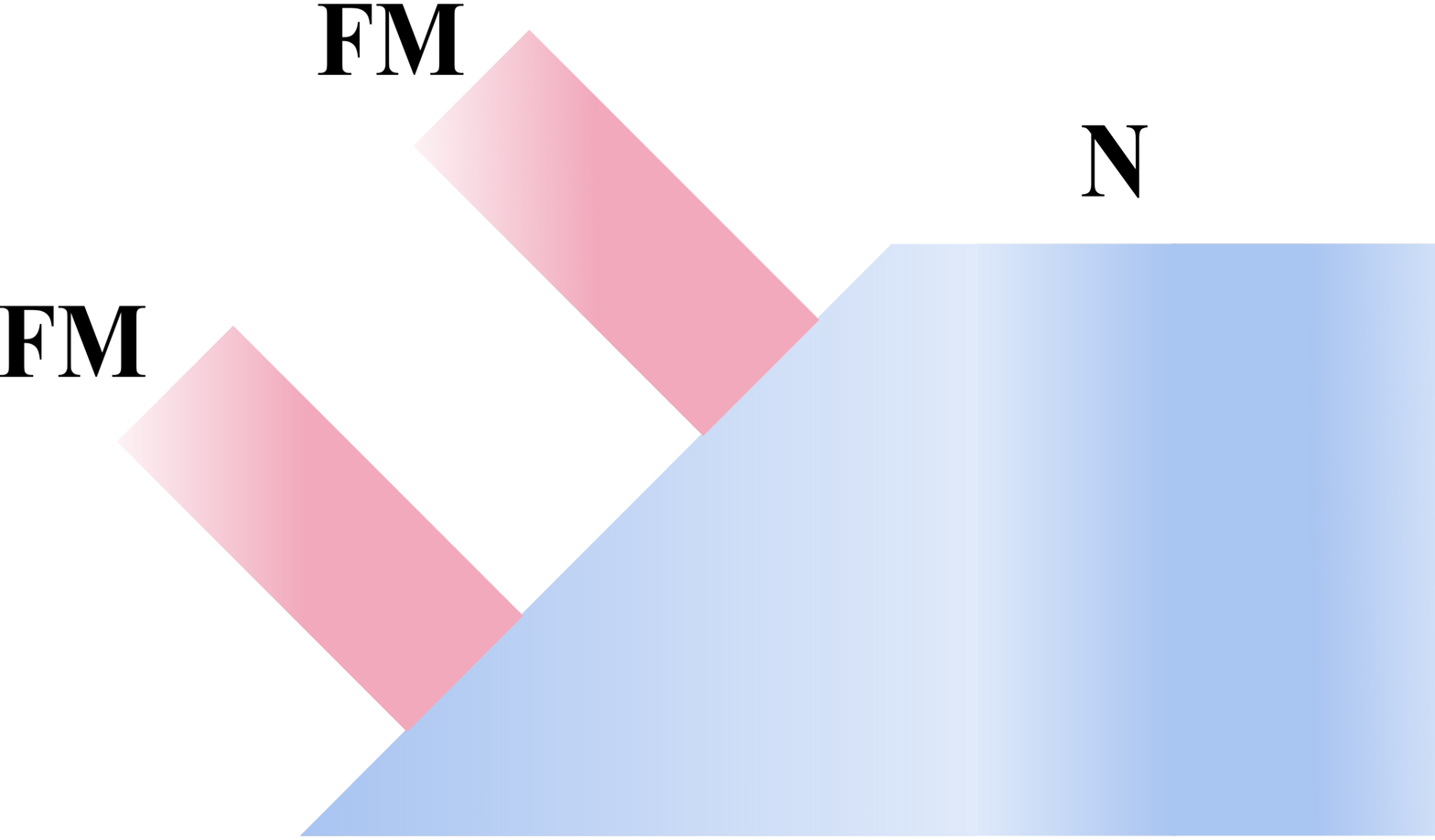}%
}
\vskip\baselineskip
\subfloat{%
\includegraphics[width=0.65\columnwidth,trim= 0.2cm 0.01cm 0.7cm 0.001cm,clip=true]{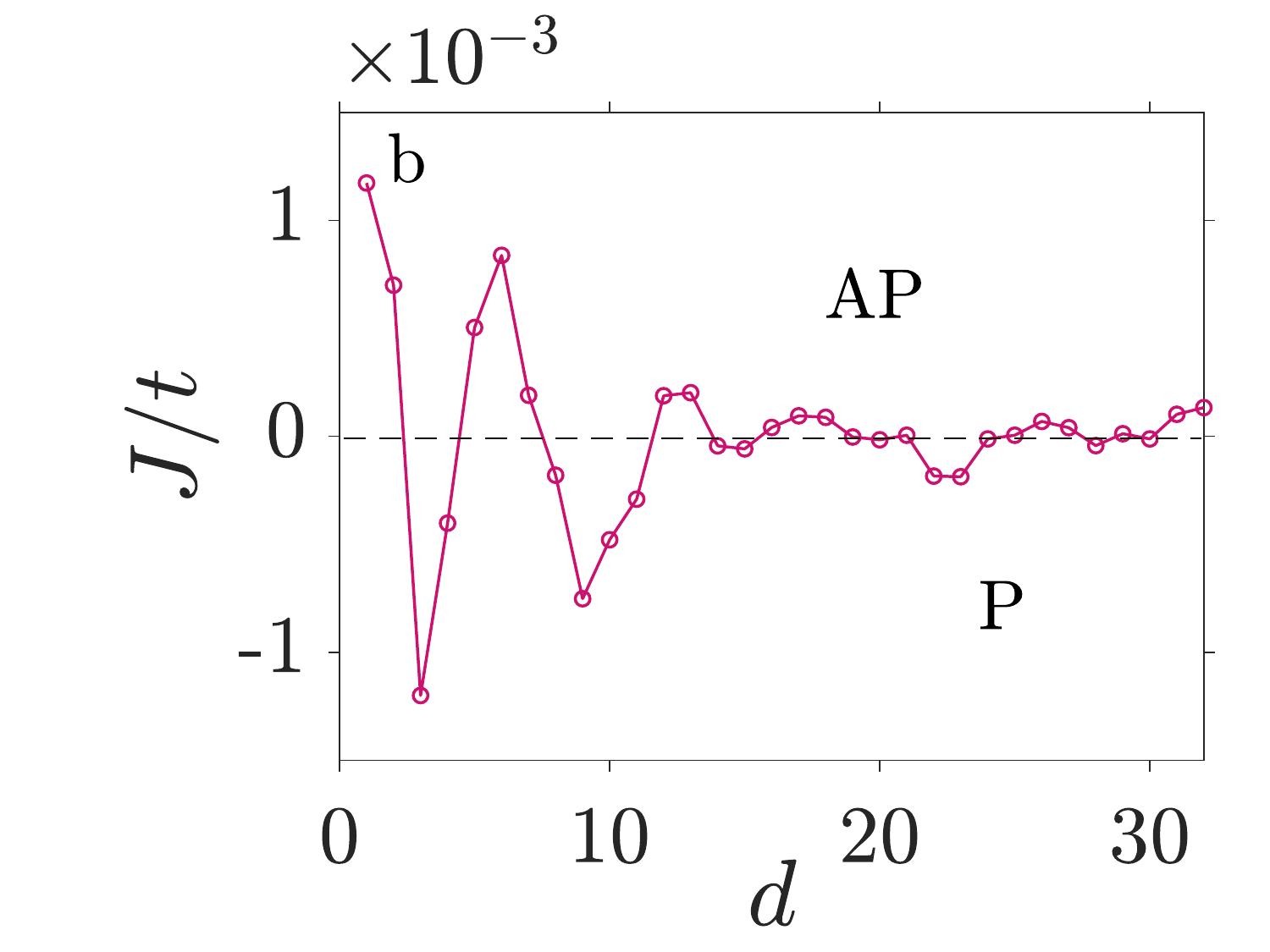}%
}
\subfloat{%
\hspace{0.0cm}
\includegraphics[width=0.35\columnwidth,trim= 0.2cm -10cm 3.0cm 0.01cm,clip=true]{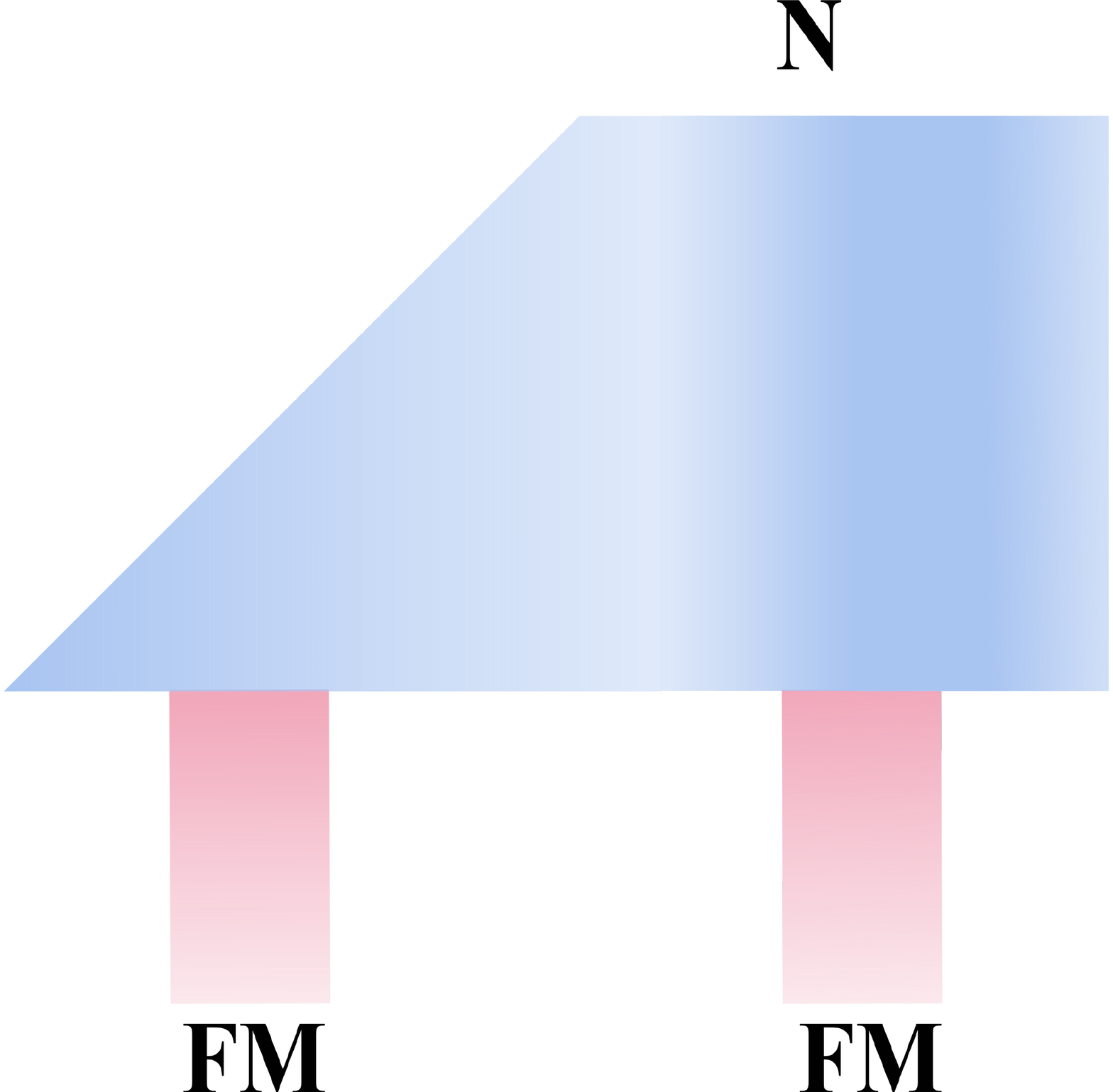}%
}
\caption{Normal metal: Indirect exchange interaction between magnetic leads connected to a diagonal (a) and a horizontal (b) edge of a normal metal, presented as a function of the distance between the leads. Here, the chemical potential in the normal metal is set to $\mu_N = 0.9t$, and the chemical potential in the ferromagnets is set to $\mu_F = 1.2t$. Further, $h_i = 2t$, $L_{xN} = 40$, $L_{yN} = 40$, $L_{xF} = 2$, $L_{yF} = 10$, and $V = U = 0$. In both subfigures, the leftmost magnet was fixed two lattice points away from the endpoint of the edge.}
\label{fig:normal-state}
\end{figure}

\subsection{Normal State}

\begin{figure}[H]
\centering
\includegraphics[width=0.5\columnwidth,trim= 0.2cm 0.01cm 25.5cm 0.001cm,clip=true]{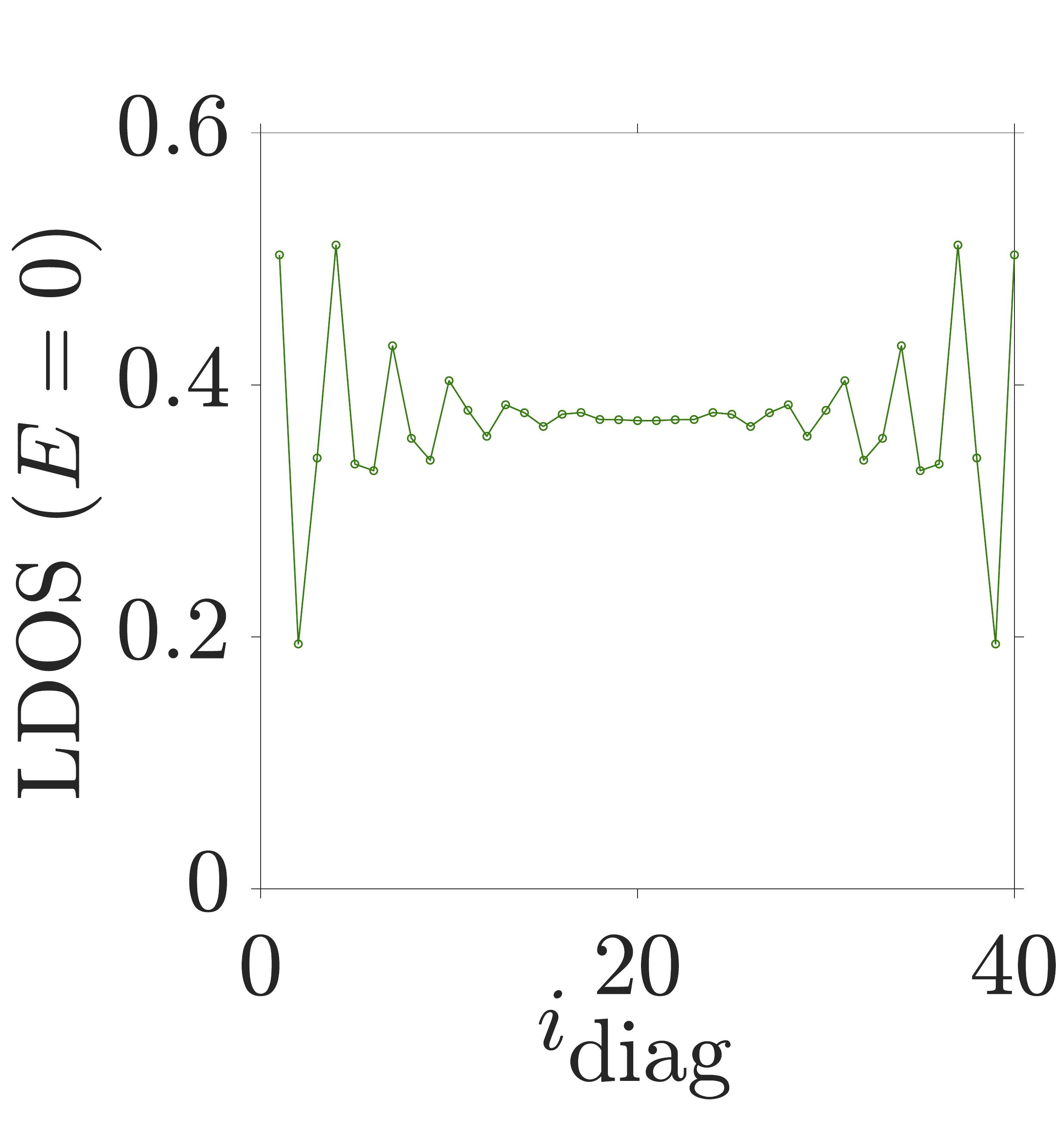}
\caption{Normal metal: Local density of states (LDOS) for $E = 0$ at the diagonal edge in the absence of magnetic contacts. The system size is $L_{xN} = 40$ and $L_{yN} = 40$, $V=U=0$, and the chemical potential in the normal metal is $0.9t$.}
\label{fig:normal_2}
\end{figure}

To put the results for the superconductors into context, we start with the case of magnetic leads connected by a normal metal ($V = U = 0$). The indirect exchange interaction $J$ is presented in Fig.\! \ref{fig:normal-state}. For the horizontal edge, the result is the expected RKKY oscillations that are damped with increasing distance between the magnets. For the diagonal edge, the results are more peculiar, showing an enhanced interaction when the electrodes are close to the endpoints of the diagonal edge. Investigating the LDOS for $E=0$ in Fig.\! \ref{fig:normal_2}, the reason becomes clear. Close to the edges of the system, the LDOS increases in magnitude and exhibits Friedel-like oscillations due to the abruptly vanishing charge density at the edge. The oscillatory and increased LDOS close to the edges correspondingly affects the RKKY interaction when the electrodes are close to the edge.

\subsection{s-wave pairing}
\begin{figure}[H]
\subfloat{%
\includegraphics[width=0.65\columnwidth,trim= 0.2cm 0.01cm 0.7cm 0.001cm,clip=true]{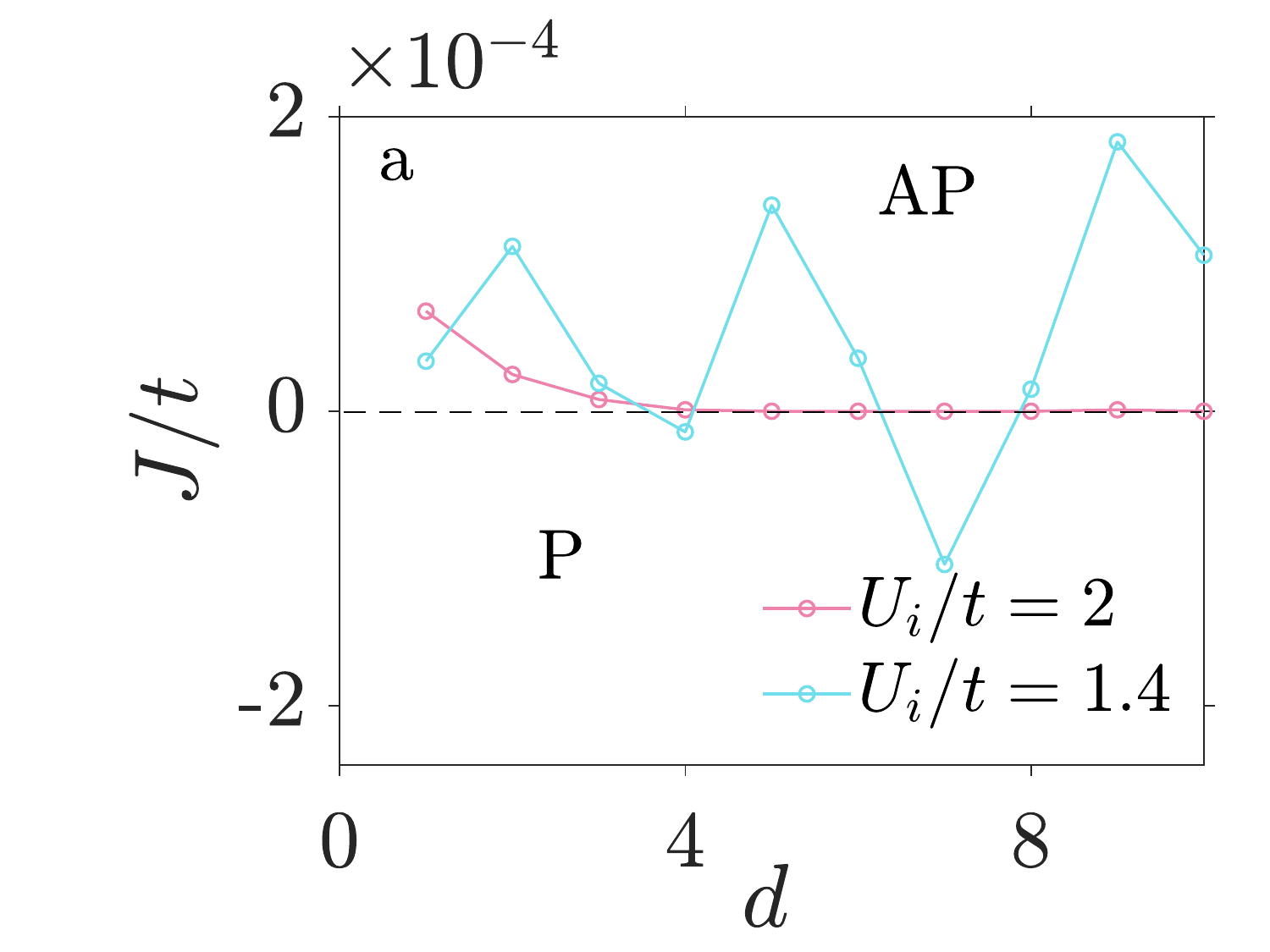}%
}
\subfloat{%
\hspace{0.1cm}
\includegraphics[width=0.35\columnwidth,trim= 0.02cm -14cm 1.0cm 0.01cm,clip=true]{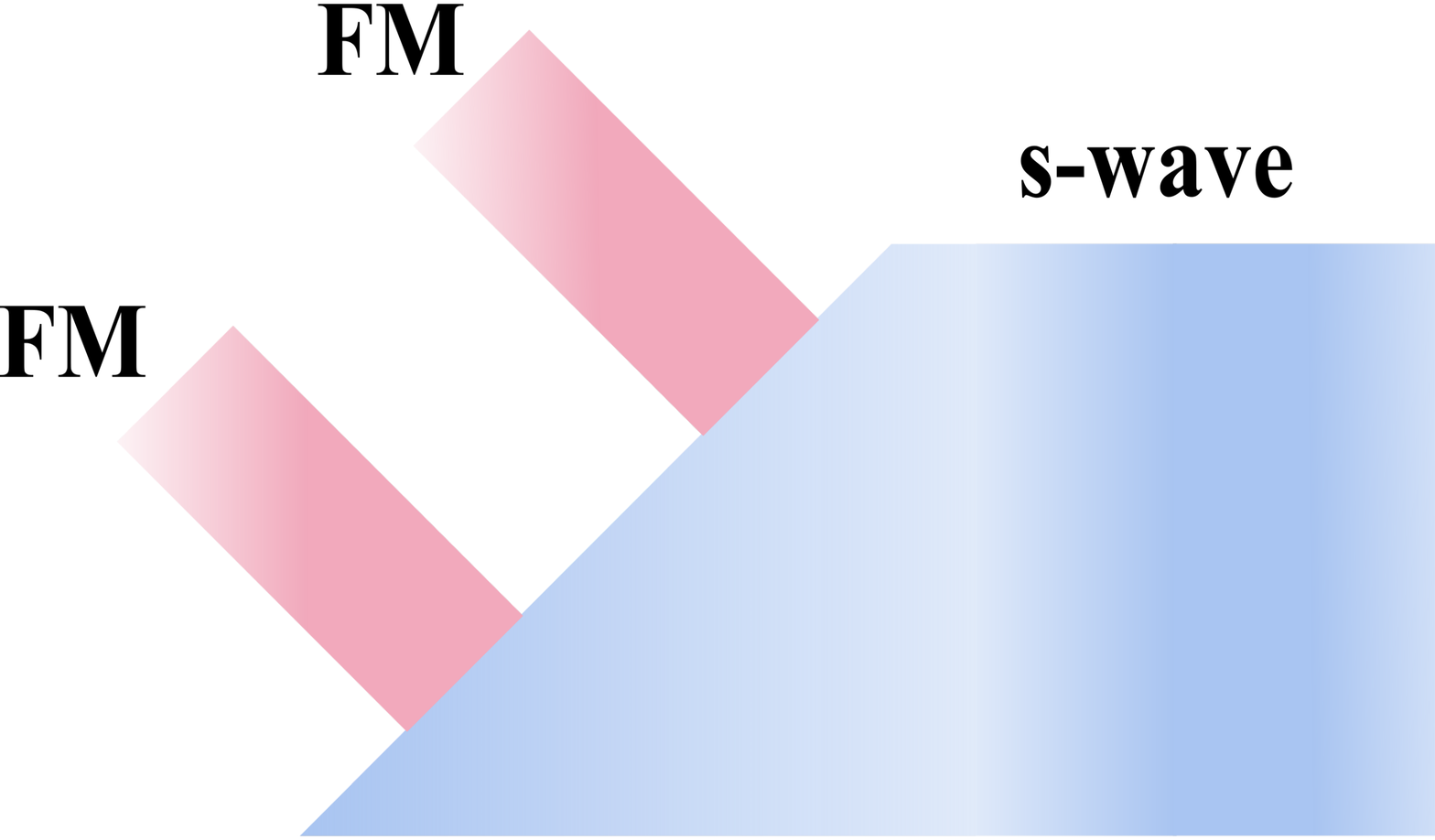}%
}
\vskip\baselineskip
\subfloat{%
\includegraphics[width=0.65\columnwidth,trim= 0.2cm 0.01cm 0.7cm 0.001cm,clip=true]{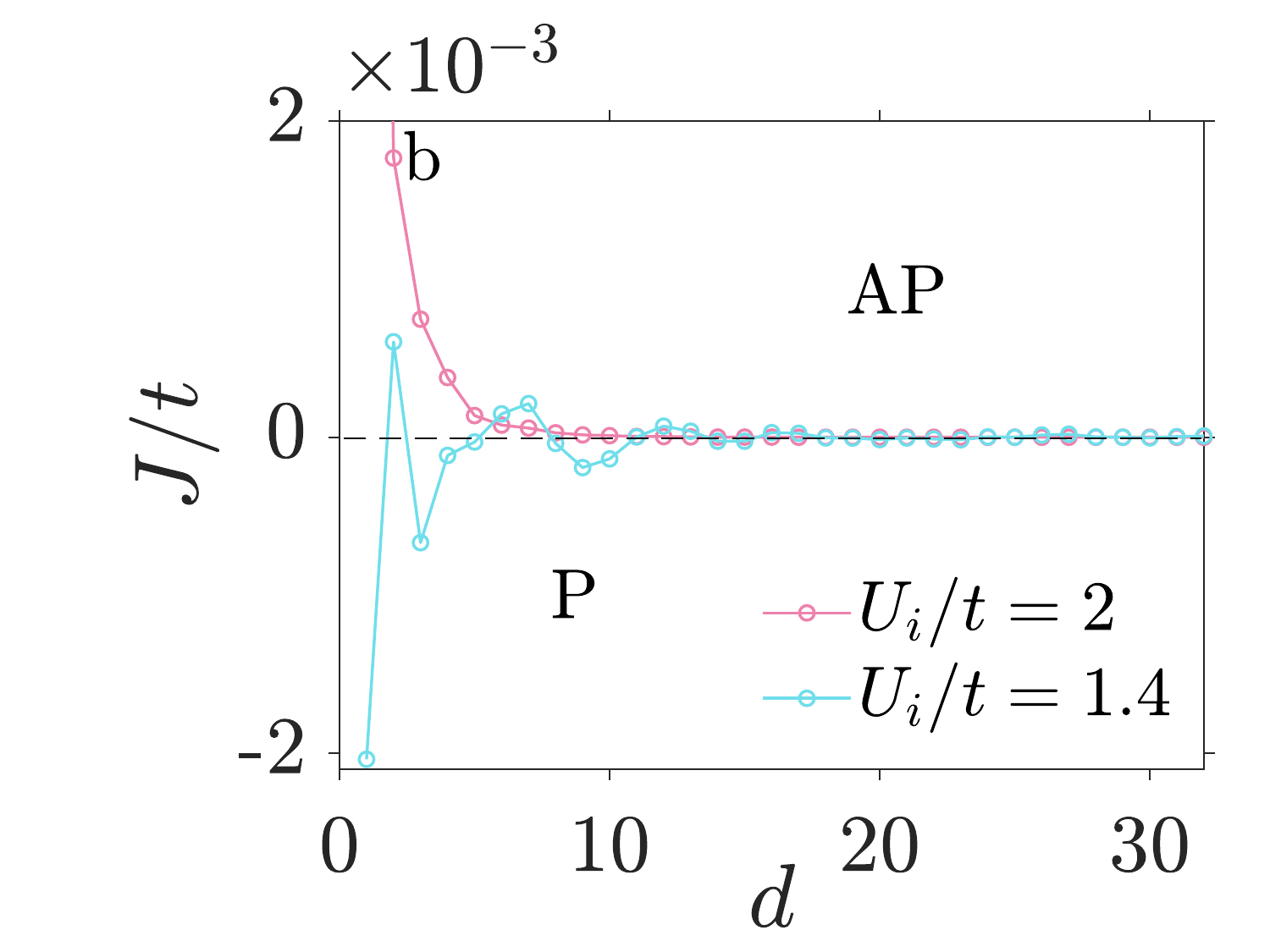}%
}
\subfloat{%
\hspace{0.0cm}
\includegraphics[width=0.35\columnwidth,trim= 0.2cm -10cm 3.0cm 0.01cm,clip=true]{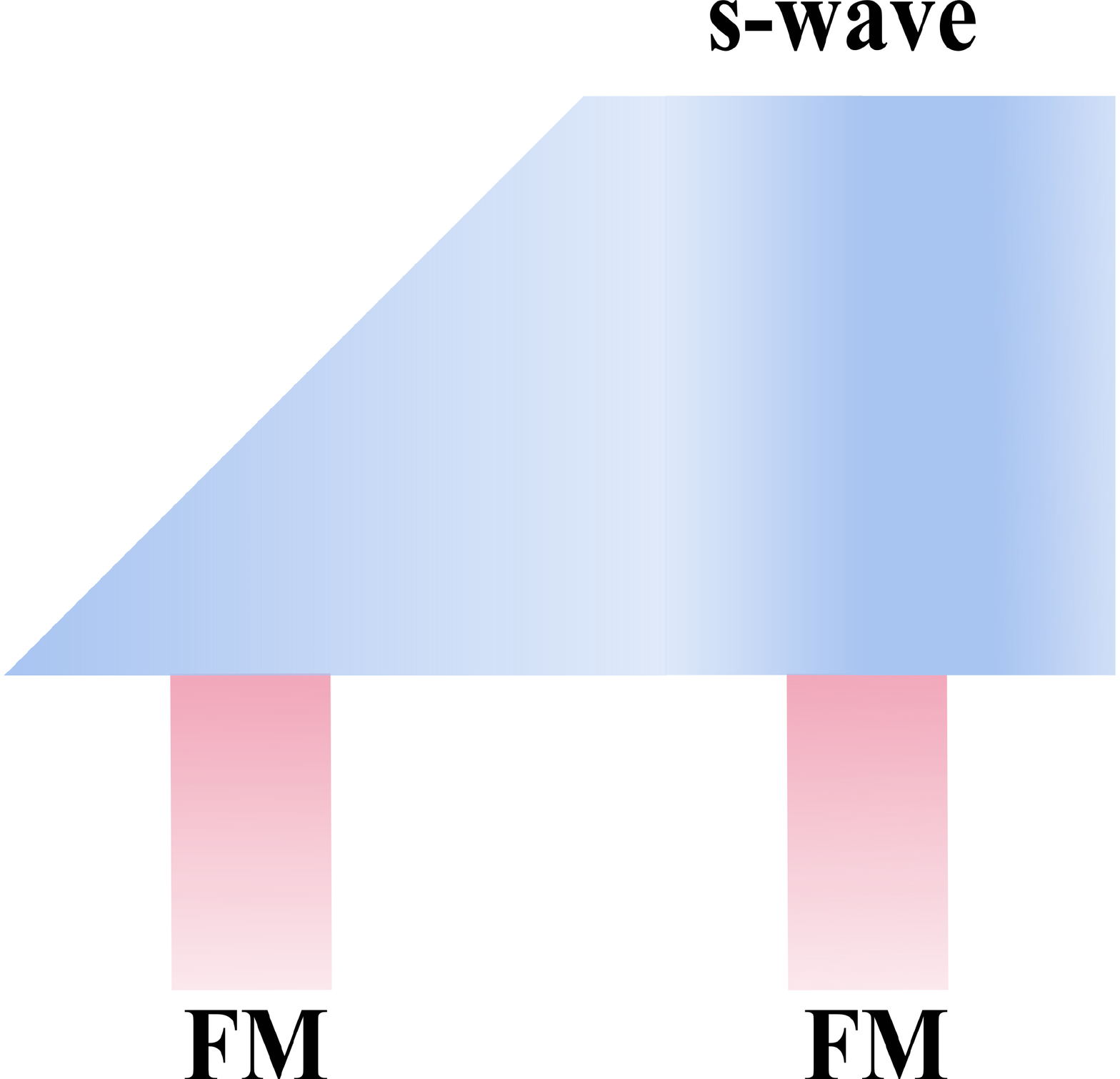}%
}
\caption{s-wave : Indirect exchange interaction between magnetic leads connected to a diagonal (a) and a horizontal (b) edge of a $s$-wave superconductor. Here $\mu_S = 0.9t$, $\mu_F = 1.2t$, $h_i = 2t$, $L_{xS} = 40$, $L_{yS} = 40$, $L_{xF} = 2$, $L_{yF} = 10$ and $V=0$. For the diagonal edge, the leftmost magnet is fixed $13$ lattice points away from the endpoint of the edge, while for the horizontal edge, the leftmost magnet is fixed $2$ lattice points away from the endpoint.}
\label{fig:s-wave}
\end{figure}

We then move on to the case of magnetic leads connected by an isotropic $s$-wave superconductor ($V = 0$). The results for the indirect exchange interaction are presented in Fig.~\ref{fig:s-wave}. In this case, there are two competing effects: the conventional RKKY interaction and the blocking of the states that can mediate the interaction due to the gap around the Fermi level in the band structure. For a weak attractive interaction $U$ in the superconductor, the RKKY interaction dominates, giving rise to an oscillating behavior. For larger $U$, the gap becomes larger and can block more of the states that can mediate the interaction between the magnets. The interaction then displays a damping behavior instead of oscillations, and an antiparallel configuration of the magnets is preferred \cite{Tussi}. For the diagonal edge, the electrodes have been kept further away from the endpoints of the edge. For a weak attractive interaction $U$, the enhanced RKKY oscillations occurring when the electrodes are close to the end-points can, however, still be observed, as explained previously. On the other hand, when the strength of the attractive interaction is increased, increasing the superconducting gap, we see that $J$ is damped to zero for sufficiently large magnet separation also for the diagonal edge. Thus, in the s-wave case, the qualitative behavior of $J$ is the
same regardless of which edge we consider.

\subsection{d-wave pairing}
\begin{figure}[H]
\subfloat{%
\includegraphics[width=0.65\columnwidth,trim= 0.2cm 0.01cm 0.7cm 0.001cm,clip=true]{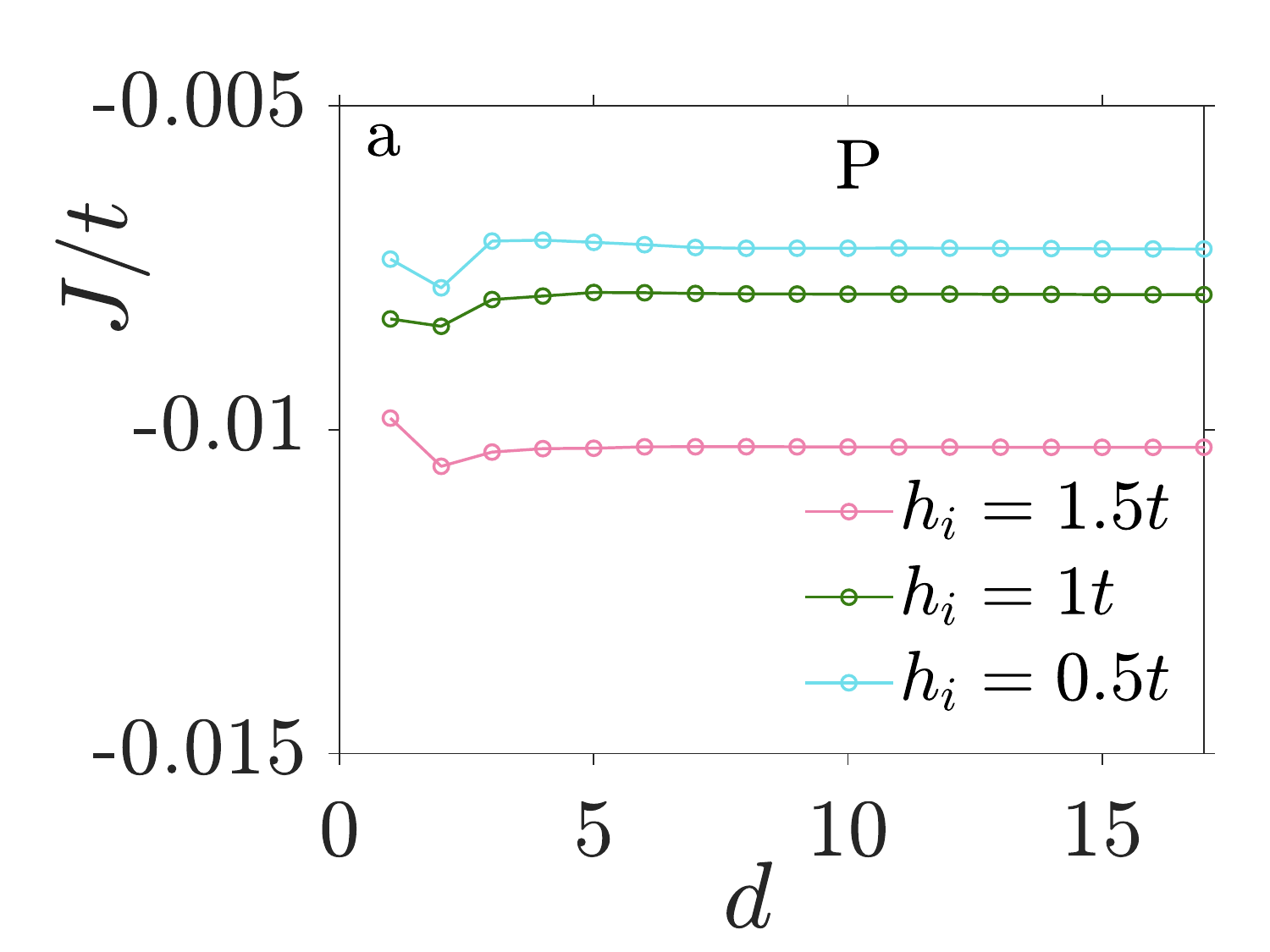}%
}
\subfloat{%
\hspace{0.1cm}
\includegraphics[width=0.35\columnwidth,trim= 0.02cm -14cm 1.0cm 0.01cm,clip=true]{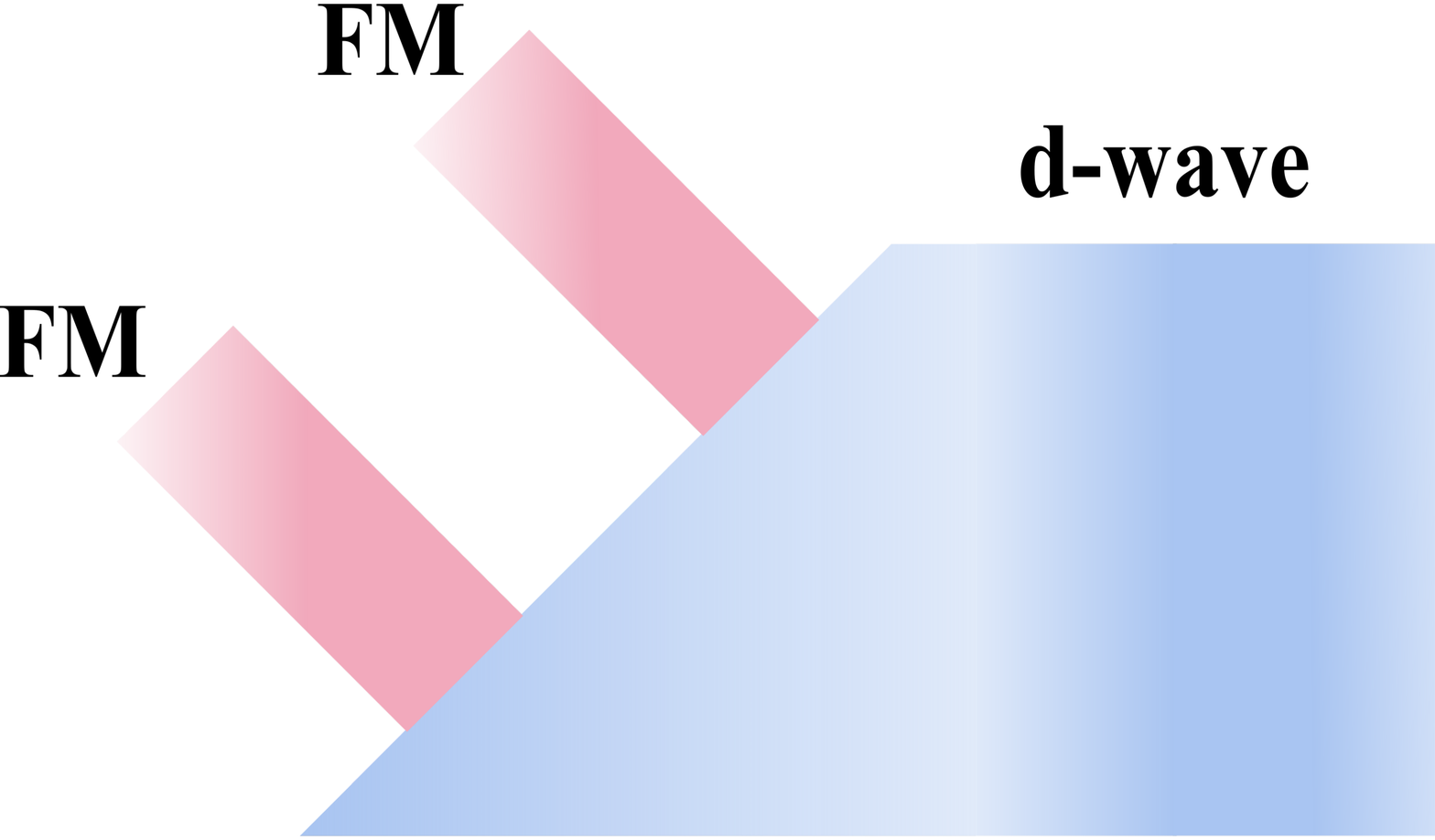}%
}
\vskip\baselineskip
\subfloat{%
\includegraphics[width=0.65\columnwidth,trim= 0.2cm 0.01cm 0.7cm 0.001cm,clip=true]{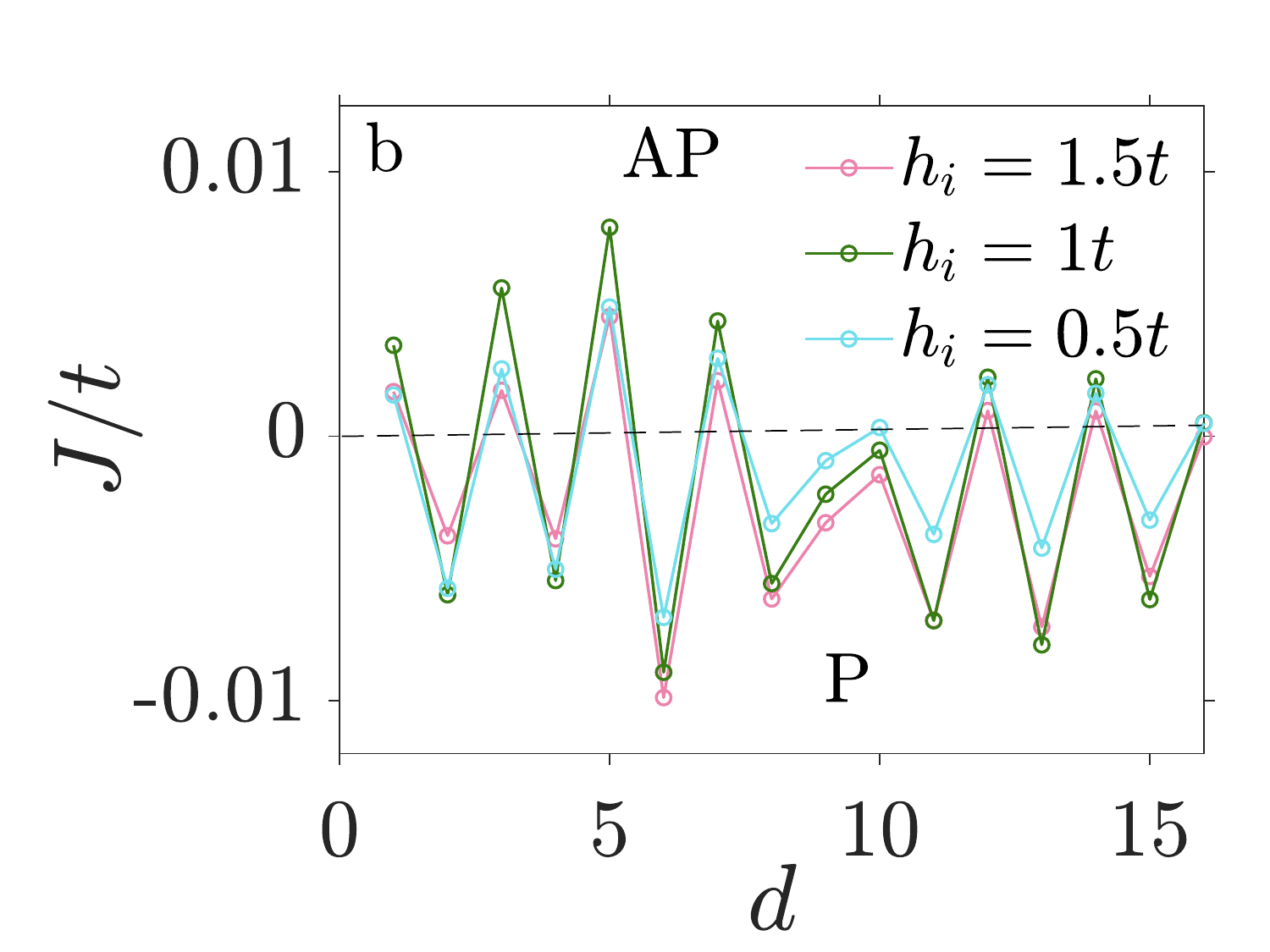}%
}
\subfloat{%
\hspace{0.0cm}
\includegraphics[width=0.35\columnwidth,trim= 0.2cm -10cm 3.0cm 0.01cm,clip=true]{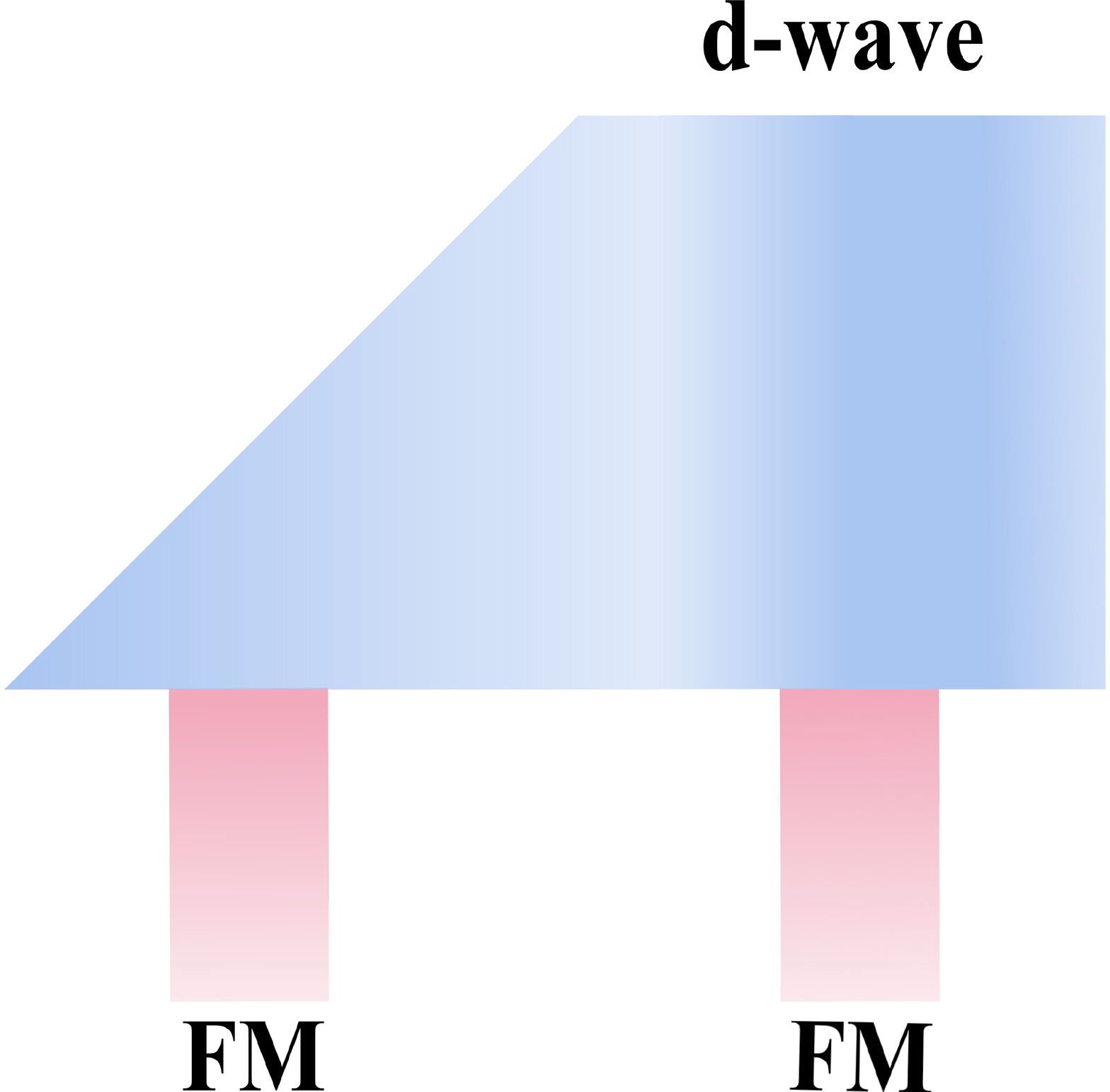}%
}
\caption{d-wave : Indirect exchange interaction between magnetic leads connected to a diagonal (a) and a horizontal (b) edge of a $d$-wave superconductor. Here $\mu_S = 0.7t$, $\mu_F = 1t$, $V/t = -1$, $L_{xF} = 2$, $L_{yF} = 10$ and $U=0$. For the diagonal edge, the leftmost magnet is fixed only $2$ lattice points away from the endpoint of the edge in order to maximize the number of data points. The $d$-wave diagonal edge results are not sensitive to how close the magnets are to the endpoints of the edge. Further $L_{xS} = 34$, $L_{yS} = 30$. For the horizontal edge, the leftmost magnet is fixed $10$ lattice points away from the endpoint, and $L_{xS} = 40$, $L_{yS} = 20$.}
\label{fig:d-wave}
\end{figure}

\begin{figure}[H]
\subfloat{%
\includegraphics[width=1\columnwidth,trim= 0.2cm 0.5cm 0.1cm 0.5cm,clip=true]{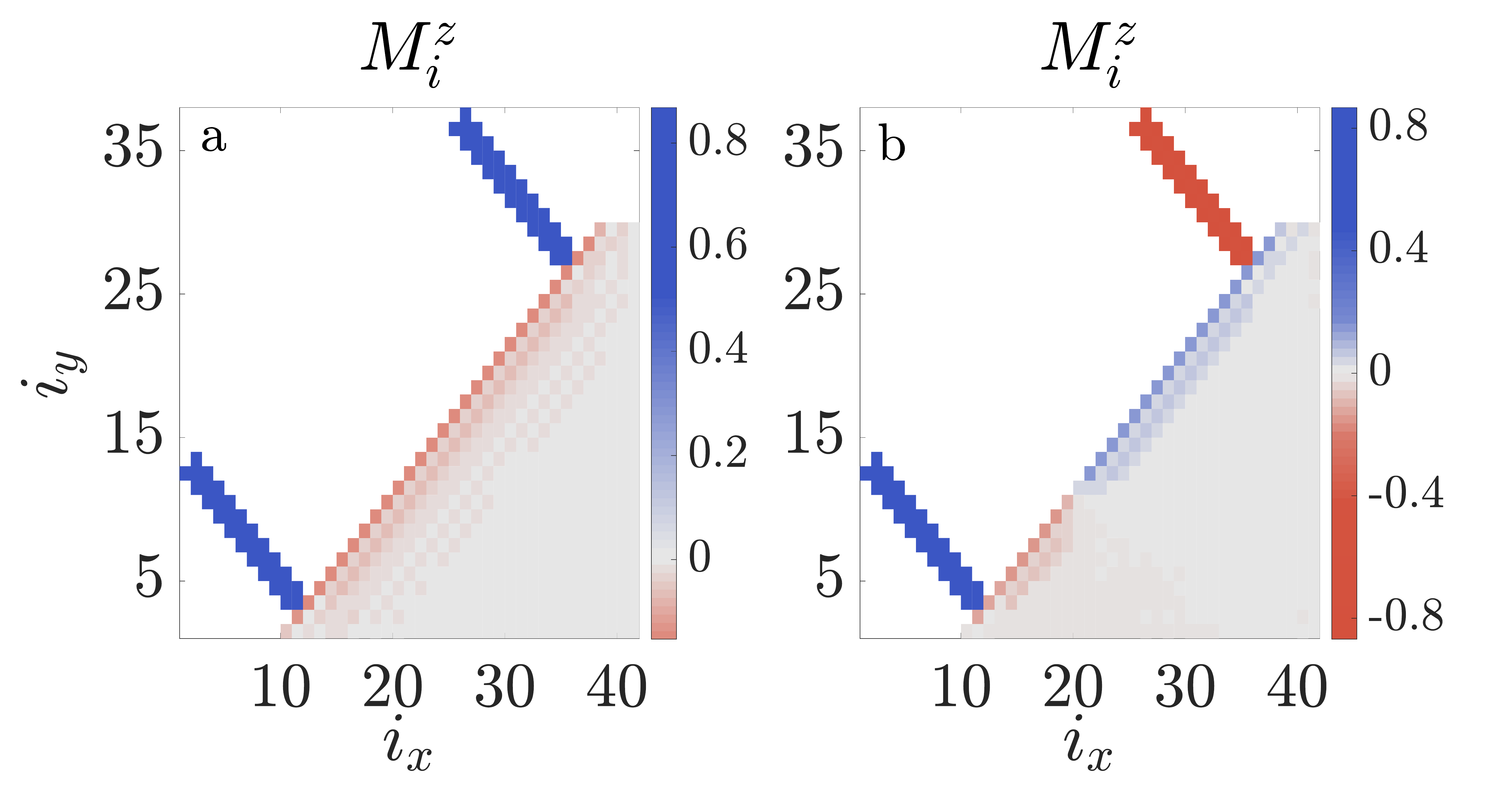}%
}

\vskip\baselineskip
\subfloat{%
\includegraphics[width=1\columnwidth,trim= 0.2cm 0.5cm 0.1cm 0.5cm,clip=true]{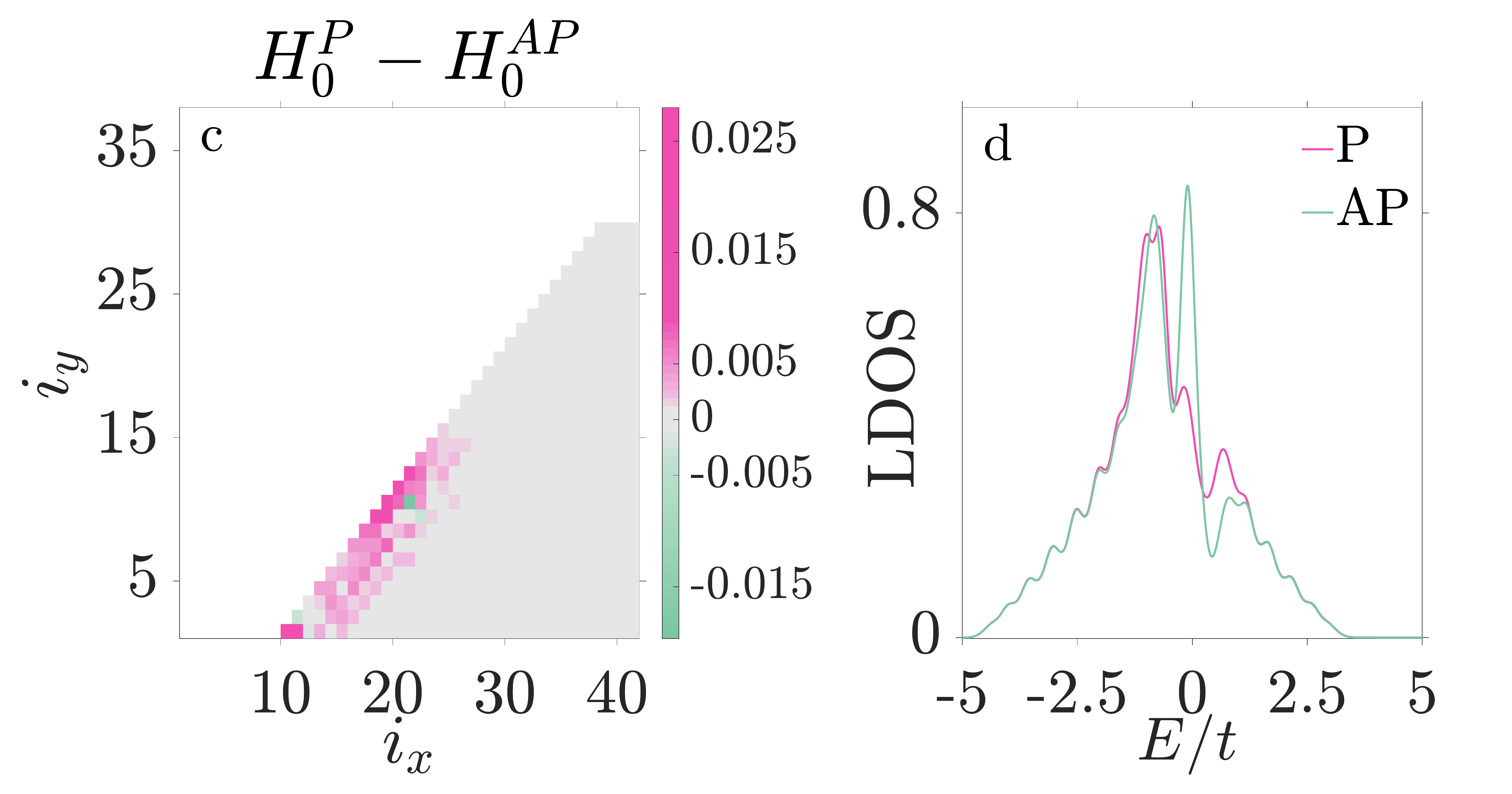}%
}
\caption{d-wave: Magnetization on each lattice site for the parallel (a) and antiparallel (b) configuration of magnets. The difference in $H_{0,i}$ between the parallel and antiparallel configurations is presented in (c). The local density of states (LDOS) for the 11th lattice site (from the left) of the diagonal edge is presented in (d), showing that the midgap states are more suppressed for the parallel magnet configuration. We have here taken the same parameters as in Fig.\! \ref{fig:d-wave} (a), apart from a larger exchange field of $h_i = 2t$ in order to more clearly show the differences between the two configurations.}
\label{fig:d-wave-reason}
\end{figure}

Finally, we consider the main result of this work, which is how the magnetic leads interact when separated by a $d$-wave superconductor ($U=0$). The results for the indirect interaction between the magnetic leads is presented in Fig.\! \ref{fig:d-wave}. For the horizontal edge, the interaction displays an oscillating behavior and varies in sign as a function of the distance between the magnetic contacts. The results for the diagonal edge, on the other hand, show a qualitatively different behavior. The system now always prefers alignment of the ferromagnets and the interaction varies little with distance. Further, increasing $h_i$ now leads to a larger difference in free energy between the parallel and antiparallel magnet configurations.\\ 
\indent The result that a parallel magnet configuration is strongly favored for the diagonal edge is surprising as one would expect that the parallel configuration induces a larger magnetization in the superconductor, which suppresses the gap and lowers
the condensation energy. A particularly large induced magnetization in the superconductor should be expected in the presence of midgap states which can give rise to a giant magnetic moment when subjected to a spin-splitting \cite{Midgap_states_Chia, Hu1999, Ting2000}. In accordance with this, we find a sizeable induced magnetization on the diagonal edge. As previously discussed in the literature, the magnetization induced in a superconductor due to proximity to a ferromagnet can be either aligned or anti-aligned with the magnetization of the ferromagnet \cite{Bergeret2005, Bergeret2005_2, Kharitonov2006}. A physical picture for the origin of an anti-aligned induced magnetization is that there are contributions from Cooper pairs where one of the
\noindent two electrons is located in the ferromagnet, aligned with the local magnetization, leaving a Cooper pair partner with opposite spin in the superconductor. In the present system the induced magnetization tends to be anti-aligned with the magnetization of the magnetic contacts, as displayed in Fig. \ref{fig:d-wave-reason} (a) and (b).

\indent The effect of the introduction of the magnets is, however, not solely to reduce the gap due to an induced effective spin-splitting in the superconductor. The induced spin-splitting also splits the midgap states away from their resonance point at zero energy, suppressing the midgap states. As the gap close to the edge to begin with is strongly suppressed by the midgap states, the effect of reducing the midgap states, causing the superconducting order parameter to recover at the edge, is stronger than the effect of the spin-splitting on the condensation energy. As the parallel configuration most effectively produces a spin-splitting in the superconductor, this configuration features the largest condensation energy, giving rise to the behavior that is observed in Fig.\! \ref{fig:d-wave} (a).\\
\indent Investigating the constant term in the Hamiltonian $H_0 = \sum_{i} H_{0,i}$, the difference between $H_{0,i}$ for the parallel and antiparallel configurations is presented in Fig.\! \ref{fig:d-wave-reason} (c). The figure shows that $H_{0}$, which is a positive quantity, is largest for the
\noindent parallel configuration, corresponding to a larger gap. In turn, this produces a larger condensation energy that lowers the free energy of the system. From the figure, it is clear that the main contribution to the difference in condensation energy between the magnetic configurations comes from the transition region where the antiparallel configuration has a reduced edge magnetization. LDOS results from this region are presented in Fig.\! \ref{fig:d-wave-reason} (d). While the AP configuration in this region has a clear midgap peak around zero energy, the midgap states for the P configuration have been split and suppressed by the induced spin-splitting.\\
\indent We close by discussing briefly experimental considerations and possible choices of materials for observation of the $d$-wave results presented in this paper. While the system sizes in the presented calculations are limited by computational considerations, the presented results are expected to be robust also for larger systems. As RKKY interaction typically decays below experimentally accessible values over short length-scales of the order of nanometers, the separation between the magnetic contacts typically needs to be kept small. This might however only apply to the RKKY dominated indirect interaction that we observe for the horizontal edge of the $d$-wave superconductor. The preference of alignment of the ferromagnets when attached to a diagonal edge of a $d$-wave superconductor is expected to also be observable for larger magnet separation as the indirect interaction in this case is not dominated by itinerant carriers, but rather arises from the parallel configuration more efficiently inducing a spin splitting, suppressing the localized midgap states. The distance the magnets can interact over is then limited by the length scale determining how far away from a magnet the midgap states still experience a spin-splitting. If the magnet separation is much larger than this decay length of the induced spin-splitting along the edge, the spin-splitting arising from each magnet decays before interacting with the spin-splitting arising from the other magnet. There is then no difference between the two magnet configurations when it comes to suppression of midgap states, and the parallel configuration is no longer favored. For an $s$-wave superconductor in proximity to a ferromagnet, the proximity-induced magnetization decays over a length scale of the superconducting coherence length \cite{Sc_proximity_Mi_Bergeret}. A natural length scale for the decay of the induced spin-splitting in the present case would then be the effective coherence length corresponding to the strongly suppressed order parameter at the edge. As the coherence length is inversely proportional to the order parameter, the magnets will then be able to interact over distances considerably larger than the bulk coherence length.\\
\indent Experimental investigation of our main finding would consist of attaching magnetic leads to a $\{110\}$ edge of a $d$-wave superconductor. The indirect interaction between the magnets can then be established by determining the energy barrier of switching between the two magnet configurations through an external magnetic field. Our prediction is that ferromagnetic alignment of the magnets will be preferred for a wide range of magnet separation distances. Possible material choices could be YBCO \cite{YBCO_1,YBCO_2, YBCO_3} for the $d$-wave superconductor featuring midgap states, and a nickel alloy like $\text{Ni}_{80}\text{Co}_{20}$ \cite{RKKY_experiment_1} for the magnetic contacts.

\section{Summary}
\label{sec:summary} 
We have investigated the indirect exchange interaction between two ferromagnetic leads connected to a superconductor as a function of the separation between the magnets, showing that the presence of zero energy surface states in a $d$-wave superconductor can qualitatively change the results. When the magnets are connected to an edge without zero energy surface states we find a normal oscillating RKKY behavior. However, when the magnets are connected to an edge featuring zero energy surface states, the strength of the magnetic exchange interaction is shifted away from zero, always favoring alignment of the magnetization in the two magnets, as the aligned configuration produces a larger superconducting condensation energy.

\section{Acknowledgements}
This work was supported by the Research Council of Norway through its Centres of Excellence funding scheme grant 262633 QuSpin.

\appendix
\section{Phase diagram}
In order to choose the parameters such that the superconductor used in the study is in a $d$-wave state, we obtain a starting point by considering a square system with continuous boundary conditions in both the $x$ and $y$ directions and no attached magnetic leads. The relevant Hamiltonian is the one in Eq.\! \eqref{eq:eq1} with $U_{\bm{i}} = V^{\prime}_{\bm{ij}} = 0$. We introduce Fourier transformations for the electron operators  $c_{\bm{i}\alpha}=\frac{1}{\sqrt{N}}\sum_{\bm{k}}e^{i \bm{k} \cdot \bm{i}}c_{\bm{k},\alpha}$ where $\bm{i} = (i_x,i_y)$ and $N$ is the number of lattice sites. After the mean-field approximation, Eq.\! \eqref{eq:eq1} then becomes

\begin{align} \label{Hamiltonian_Sc_k_space}
\begin{aligned}
H^{SC}= &  -\sum_{\langle \bm{i},\bm{j} \rangle,\alpha} t c_{\bm{i}\alpha}^{\dagger}c_{\bm{j}\alpha}-\sum_{\bm{i},\alpha} \mu_{\bm{i}} n_{\bm{i}\alpha} \\
        &  + \!\sum_{\bm{i}, \alpha \neq \alpha^{\prime}} V \Big[ n_{\bm{i}\alpha} n_{\bm{i}+\hat{x},\alpha^{\prime}} + n_{\bm{i}\alpha} n_{\bm{i}-\hat{x},\alpha^{\prime}} \\
        &  + n_{\bm{i}\alpha} n_{\bm{i}+\hat{y},\alpha^{\prime}} + n_{\bm{i}\alpha} n_{\bm{i}-\hat{y},\alpha^{\prime}} \Big] = \sum_{\bm{k},\sigma} \zeta_{\bm{k}} c_{\bm{k},\sigma}^{\dagger} c_{\bm{k},\sigma} \\
        &  + \sum_{\bm{k}} \Big [ (\Gamma_{\bm{k}})^{\dagger} c_{\bm{k} \downarrow}^{\dagger} c_{-\bm{k} \uparrow} ^{\dagger} + \Upsilon_{\bm{k}} c_{\bm{k} \uparrow} c_{-\bm{k} \downarrow} \big ] + H_0^{SC}.
\end{aligned}
\end{align}
\noindent Here $H_0^{SC} = -2NV (|F^{\hat{x} +}|^2 + |F^{\hat{x} -}|^2 + |F^{\hat{y} +}|^2 + |F^{\hat{y} -}|^2)$, 

\begin{align} \label{F_amplitudes}
\begin{aligned}
F^{x \pm} = & \frac{1}{N}\sum_{\bm{k}} e^{\mp i \bm{k} \cdot \hat{x}} \langle c_{\bm{k},\uparrow} c_{- \bm{k},\downarrow} \rangle,
\\
F^{y \pm} = & \frac{1}{N} \sum_{\bm{k}} e^{\mp i\bm{k} \cdot \hat{y}} \langle c_{\bm{k},\uparrow} c_{- \bm{k},\downarrow} \rangle ,
\end{aligned}
\end{align}

and we have defined,
\begin{equation}\label{eq:eqA6}
\begin{array}{l@{\qquad}l}
\Upsilon_{\bm{k}} = 2V(e^{-i \bm{k} \cdot \hat{x}} (F^{\hat{x} +})^{\dagger} + e^{i \bm{k} \cdot \hat{x}} (F^{\hat{x} -})^{\dagger} +
\\\\
\hspace{1.4 cm} e^{-i \bm{k} \cdot \hat{y}} (F^{\hat{y} +})^{\dagger} + e^{i \bm{k} \cdot \hat{y}} (F^{\hat{y} -})^{\dagger} ),
\\\\
\Gamma_{\bm{k}} = 2V(e^{i \bm{k} \cdot \hat{x}} (F^{\hat{x} +})^{\dagger} + e^{-i \bm{k} \cdot \hat{x}} (F^{\hat{x} -})^{\dagger} +
\\\\
\hspace{1.4 cm} e^{i \bm{k} \cdot \hat{y}} (F^{\hat{y} +})^{\dagger} + e^{-i \bm{k} \cdot \hat{y}} (F^{\hat{y} -})^{\dagger} ),
\\\\
\varepsilon_{\bm{k}} = -2t[ \cos(\bm{k} \cdot \hat{x}) + \cos(\bm{k} \cdot \hat{y}) ] - \mu. 
\end{array}
\end{equation}
Further, $t = t_{\bm{ij}}$ and $V = V_{\bm{ij}}$.\\
\indent Following the BdG method \cite{deGennes1999}, we consider the following basis in order to diagonalize the Hamiltonian  
\begin{equation}\label{eq:eqA2}
   B_{\bm{k}}^{\dagger}=
  \left[ {\begin{array}{cccc}
    c_{\bm{k}\uparrow}^{\dagger} & c_{\bm{k}\downarrow}^{\dagger} & c_{-\bm{k}\uparrow} & c_{-\bm{k}\downarrow} 
  \end{array} } \right].
\end{equation}
Then full Hamiltonian can be written as $H=H_0 + \frac{1}{2} \sum_{\bm{k}} B_{\bm{k}}^{\dagger} H_{\bm{k}} B_{\bm{k}}$, where $H_0 = H_0^{SC} + \sum_{\bm{k}}\varepsilon_{\bm{k}}$ and $H_{\bm{k}}$ is

\begin{equation}\label{eq:eqA5}
   H_{\bm{k}}=
  \left[ {\begin{array}{cccc}
    \varepsilon_{\bm{k}}  & 0 & 0 & - (\Upsilon_{\bm{k}})^{\dagger} \\\\

    0 & \varepsilon_{\bm{k}} & (\Gamma_{\bm{k}})^{\dagger} & 0\\\\

    0 & \Gamma_{\bm{k}} & -\varepsilon_{\bm{k}} & 0\\\\
    
     - \Upsilon_{\bm{k}} & 0 & 0 & -\varepsilon_{\bm{k}}
  \end{array} } \right].
\end{equation}
Using the unitary matrix $P_{\bm{k}}$ the diagonalized form of the Hamiltonian will be $H^{SC} = H_0 + \frac{1}{2} \sum_{\bm{k}} B_{\bm{k}}^{\dagger} P_{\bm{k}}^{\dagger} P_{\bm{k}} H_{\bm{k}} P_{\bm{k}}^{\dagger} P_{\bm{k}} B_{\bm{k}} = H_0 + \frac{1}{2}  \sum_{\bm{k}} \tilde{B_{\bm{k}}}^{\dagger} \tilde{H_{\bm{k}}} \tilde{B_{\bm{k}}} = H_0 - \frac{1}{2} \sum_{\bm{k},\sigma} E_{\bm{k},\sigma} + \sum_{\bm{k}, \sigma} E_{\bm{k},\sigma} \gamma_{\bm{k} \sigma}^{\dagger} \gamma_{\bm{k} \sigma}$. The relationship between the normal electron operators and the quasiparticle operators is then

\begin{equation}
  \left[ {\begin{array}{ccccc}
    \upsilon_{\bm{k}, \uparrow} & \upsilon_{\bm{k}, \downarrow} & \omega_{-\bm{k}, \uparrow}^{\ast} & \omega_{-\bm{k}, \downarrow}^{\ast} \\
    \nu_{\bm{k}, \uparrow} & \nu_{\bm{k}, \downarrow} & \chi_{-\bm{k}, \uparrow}^{\ast} & \chi_{-\bm{k}, \downarrow}^{\ast} \\
    \omega_{\bm{k}, \uparrow} & \omega_{\bm{k}, \downarrow} & \upsilon_{-\bm{k}, \uparrow}^{\ast} & \upsilon_{-\bm{k}, \downarrow}^{\ast}\\
    \chi_{\bm{k}, \uparrow} & \chi_{\bm{k}, \downarrow} & \nu_{-\bm{k},\uparrow}^{\ast} & \nu_{-\bm{k},\downarrow}^{\ast} \\
  \end{array} } \right]
   \left[ {\begin{array}{c}
    \gamma_{\bm{k}\uparrow} \\
    \gamma_{\bm{k}\downarrow}  \\
    \gamma_{-\bm{k}\uparrow}^{\dagger}\\
    \gamma_{-\bm{k}\downarrow}^{\dagger}\\
  \end{array} } \right] =
   \left[ {\begin{array}{c}
    c_{\bm{k}\uparrow} \\
    c_{\bm{k}\downarrow}  \\
    c_{-\bm{k}\uparrow}^{\dagger}\\
    c_{-\bm{k}\downarrow}^{\dagger}\\
  \end{array} } \right],
\end{equation}
where the columns are the eigenvectors of $H_{\bm{k}}$. The pairing amplitudes can then be expressed as 

\begin{align} \label{F_amplitudes}
    \begin{aligned}
        F^{x \pm} =  \frac{1}{N}\sum_{\bm{k}, \sigma}  \Big [& e^{\mp i \bm{k} \cdot \hat{x}} \upsilon_{\bm{k},\sigma}  \chi_{\bm{k},\sigma}^{\ast}  \big( 1 - f( E_{\bm{k},\sigma}) \big)  \\
        &+ e^{\pm i \bm{k} \cdot \hat{x}} \omega_{\bm{k},\sigma}^{\ast} \nu_{\bm{k},\sigma} f( E_{\bm{k},\sigma})  \Big ],
        \\
        F^{y \pm} =   \frac{1}{N} \sum_{\bm{k}, \sigma} \Big [& e^{\mp i\bm{k} \cdot \hat{y}} \upsilon_{\bm{k},\sigma}  \chi_{\bm{k},\sigma}^{\ast}  \big( 1 - f( E_{\bm{k},\sigma}) \big) \\
        &+ e^{\pm i\bm{k} \cdot \hat{y}} \omega_{\bm{k},\sigma}^{\ast} \nu_{\bm{k},\sigma} f( E_{\bm{k},\sigma})  \Big ].
    \end{aligned}
\end{align}

\noindent Finally, the free energy of the system is 

\begin{equation} \label{Free_energy_k_space}
    F = H_0 - \frac{1}{2} \sum_{\bm{k},\sigma} E_{ \bm{k},\sigma} -\frac{1}{\beta} \sum _{\bm{k}, \sigma} \text{ln} (1 + e^{-\beta E_{\bm{k},\sigma}}). 
\end{equation}

\renewcommand{\arraystretch}{1.1}
\setlength{\tabcolsep}{3pt}

\begin{table}[tbp]
\centering 
\caption{Sets of initial values.}
\label{tab:table1}
\begin{tabular}{llllll}
\hline
& \,\, & $F^{\hat{x} +}$ & $F^{\hat{x} -}$ & $F^{\hat{y} +}$ & $F^{\hat{y} -}$\\
\hline\\
$d$-wave & & 1 &  \,\,1 & -1 & -1 \\\\
$s$-wave extended & & 1 &  \,\,1 & \,\,1 & \,\,1\\\\
$p_x + ip_y$ & & 1 & -1 & \,\,$i$ & \,\,-$i$ \\\\
Normal state & & 0 & \,\,0 & \,\,0 & \,\,0 \\\\
\hline
\end{tabular}
\end{table}

\begin{figure}[H]
\centering
\includegraphics[width=0.95\columnwidth,trim= 0.005cm 0.01cm 1.0cm 0.01cm,clip=true]{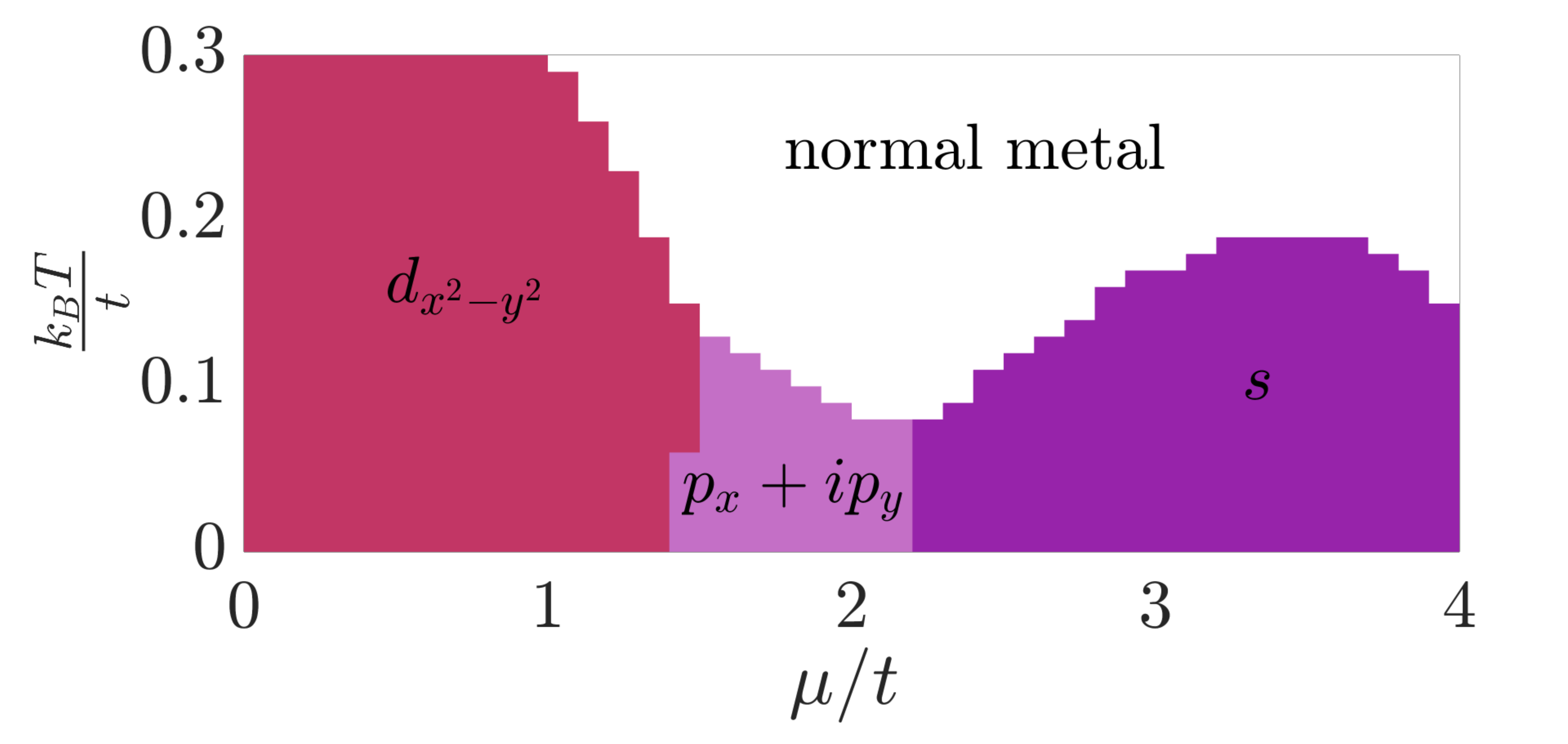}%
\caption{Phase diagram for the tight-binding Hamiltonian with attractive nearest neighbor interaction between opposite spins $V = - 1t$. Here, $T$ is the temperature, $k_B$ is the Boltzmann constant, $\mu$ is the chemical potential, and $t$ is the hopping amplitude. }
\label{fig:phase-plot}
\end{figure}

\indent For different values of chemical potential and temperature, we then solve the self-consistent equations for the pairing amplitudes through iteration, using the different sets of initial values listed in table \ref{tab:table1}. We then compare the resulting free energies (Eq.\! ~\eqref{Free_energy_k_space}) and determine the favored phase of the system. The phase diagram is presented in Fig.\! \ref{fig:phase-plot}. The choices for the initial values are determined by the expressions for the gap functions \cite{Gap_equations}
\begin{align}
\begin{aligned}
    &\Delta_d     = (V/4) (F^{\hat{x} +} + F^{\hat{x} -} - F^{\hat{y} +} - F^{\hat{y} -}),\\
    &\Delta_{s}   = (V/4) (F^{\hat{x} +} + F^{\hat{x} -} + F^{\hat{y} +} + F^{\hat{y} -}),\\
    &\Delta_{p_x} = (V/2) (F^{\hat{x} +} - F^{\hat{x} -}),\\
    &\Delta_{p_y} = (V/2) (F^{\hat{y} +} - F^{\hat{y} -}).
\end{aligned}
\end{align}


\bibliographystyle{apsrev4-1}
\addcontentsline{toc}{chapter}{\bibname}
\bibliography{Refs}

\end{document}